
\magnification=\magstep1
\hsize = 32pc
\vsize = 46pc
\baselineskip=14pt
\tolerance 2000
\parskip=4pt

\font\grand=cmbx10 at 14.4truept
\font\grand=cmbx10 at 14.4truept

\font \ninerm                 = cmr9

\pageno=0
\def\folio{
\ifnum\pageno<1 \footline{\hfil} \else\number\pageno \fi}


\def\Sei{{1}}
\def\GPP{{2}}
\def\Mar{{3}} 
\def\Bal{{4}} 
\def\WZ{{5}}
\def\FT{{6}}
\def\MW{{7}}
\def\LP{{8}}
\def\Ki{{9}}
\def\Seg{{10}}
\def\Wi{{11}}
\def\OK{{12}}
\def\GRS{{13}}
\def\G{{14}}
\def\Fad{{15}}
\def\Ald{{16}}
\def\Liu{{17}}
\def\THE{{18}}
\def\OTW{{19}}
\def\GERC{{20}}
\def\GJ{{21}}
\def\Ful{{22}}
\def\KT{{23}}
\def\RY{{24}}
\def\BPT{{25}}
\def\KS{{26}}


\def\pa{\partial}
\def\R{{\bf R}}
\def\L{{\cal L}}
\def\rA{{\rm A}}

\def\pa{\partial}

\def\tr{{\rm tr}}
\def\ln{\,{\rm ln}\,}
\def\cR{{\cal R}}
\def\cF{{\cal F}}
 

\rightline{BONN-TH-97-02}
\rightline{ITP Budapest Report No.\ 526}

\vskip 1.0truecm

\centerline{\grand Coadjoint orbits of the Virasoro algebra} 

\centerline{\grand and the global Liouville equation}

\vskip 1.0truecm

\centerline{J.\ Balog${}^{1,}$\footnote{${}^\dagger$}{\ninerm On 
leave from the Research Institute for
Particle and Nuclear Physics, Budapest, Hungary},  
L.\  Feh\'er${}^{2}$ and L. Palla${}^{3}$}

\medskip

{\baselineskip=12pt 
\centerline{\it 
${}^1$Laboratoire de Math.\ et Physique Th\' eorique,
CNRS UPRES-A 6083}

\centerline{\it Universit\' e de Tours,
Parc de Grandmont, F-37200 Tours, France} 
}

\medskip
\centerline{\it 
${}^{2}$Physikalisches Institut der Universit\" at Bonn,
Nussallee 12,  D-53115 Bonn, Germany}

\medskip

{\baselineskip = 12pt
\centerline{\it ${}^3$Institute for Theoretical Physics,
Roland E\" otv\" os University}

\centerline{\it H-1088 Budapest, Puskin u 5-7, Hungary}
}

\vskip 1.0 truecm 
\centerline{\bf Abstract}
\medskip

The classification of the coadjoint orbits of the Virasoro algebra
is reviewed and is then applied to analyze the so-called global Liouville 
equation. The review is self-contained, elementary and is tailor-made for 
the application. It is well-known that the Liouville equation for a smooth, 
real field $\varphi$ under periodic boundary condition is a reduction  of 
the $SL(2,\R)$ WZNW model on the cylinder, where the WZNW field 
$g\in SL(2,\R)$ is restricted to be Gauss decomposable. If one drops this 
restriction,  the Hamiltonian reduction yields, for the field 
$Q=\kappa g_{22}$ where $\kappa\neq 0$ is a constant, what we call the 
global Liouville equation. Corresponding  to the winding number of the 
$SL(2,\R)$ WZNW model there is a topological invariant in the reduced 
theory, given by the number of zeros of $Q$ over a period. By the 
substitution $Q=\pm\exp( - \varphi/2)$, the Liouville theory for a smooth 
$\varphi$ is recovered in the trivial topological sector. The nontrivial 
topological sectors can be viewed as singular sectors of the Liouville 
theory that contain blowing-up solutions in terms of $\varphi$.
Since the global Liouville equation is conformally invariant, its solutions
can be described by explicitly listing  those  solutions for which the 
stress-energy tensor belongs to a set of representatives of the Virasoro 
coadjoint orbits chosen by convention. This direct method permits to study 
the `coadjoint orbit content'  of the topological sectors as well as the 
behaviour of the energy in the sectors.  The analysis  confirms that the 
trivial topological sector contains special orbits with hyperbolic monodromy
and shows that the energy is bounded from below in this  sector only.
 
\vfill\eject

\centerline{\bf 1.~Introduction}
\medskip

The Liouville equation plays an important role in various areas of
theoretical physics (for a review, see e.g.\ Ref.~[\Sei]).
Our motivation for the present work was to  
explore the space of solutions of the evolution equation  
$$
Q\pa_+\pa_- Q - \pa_+ Q \pa_- Q =1,
\qquad 
x^\pm = {\tau \pm \sigma\over 2}, \quad 
\pa_{\pm} = (\pa_\tau \pm \pa_\sigma),
\eqno(1.1)$$
for a {\it smooth},
{\it real} field\footnote{${}^{a}$}{\ninerm In this paper {\it smooth} 
means infinitely many times differentiable, i.e, $C^\infty$.} 
$Q$ under {\it periodic} boundary
condition in $\sigma$.
We call this equation the {\it global} Liouville equation, since it 
becomes the Liouville equation
$$
\pa_+ \pa_- \varphi +2 e^\varphi =0
\eqno(1.2)$$
upon the substitution $Q= \pm \exp(-\varphi/2)$, which is 
valid only {\it locally},  if $Q$ has no zeros.
Notice that the number of zeros of $Q$ over a period in $\sigma$ is a 
{\it topological invariant} for the global Liouville theory.
It cannot be changed by a smooth deformation of $Q$, since (1.1)
implies that $Q$  cannot have any double zero in $\sigma$. 
The number of zeros, $N$,  is even as $Q$ is periodic.
The Liouville equation (1.2) for a smooth $\varphi$ is 
recovered in the trivial, $N=0$  
 topological sector of (1.1), and the other
sectors may be thought of as singular sectors of the Liouville theory,
where the solutions blow up in terms of $\varphi$, but remain smooth
in terms of $Q$.
The smoothness of $Q$ is in fact equivalent to the smoothness 
of the conformally improved Liouville stress-energy tensor.
One of the main questions about the 
model is how the energy behaves
in the various topological sectors. 
It will turn out in the end that the energy is bounded from 
below in the trivial topological sector only.

The singular solutions of the Liouville equation were  
studied previously [\GPP,\Mar]  
using methods different from those that will be used here.
Equation (1.1) was apparently first introduced in the second article of
Ref.~[\Mar].  
We were led to the global Liouville equation by the 
earlier  treatment [\Bal] 
of the Liouville theory as a reduced   Wess-Zumino-Novikov-Witten 
(WZNW) model [\WZ]  
on the cylinder for the group $SL(2,\R)$.
In that case, the Liouville equation arises if one further restricts 
the group valued field $g\in SL(2,\R)$ to vary in the neighbourhood of the
identity  consisting of Gauss decomposable matrices.
Dropping this restriction, one obtains precisely the global Liouville
equation for $Q= \kappa g_{22}$,
where $\kappa$ is the level parameter of the WZNW model. 
Hence, studying  the solutions of (1.1) helps 
to explore the global structure of the phase space of the reduced 
WZNW model [\FT]. 
The topological sectors of the global
Liouville theory correspond to the 
topological sectors of the WZNW model,
which are labelled  by the winding number  
of the field $g: S^1\times \R\rightarrow SL(2, \R)$. 
The constraints of the reduction 
enforce that the  winding number equals 
${\rm sign\,}\kappa$-times half the number of zeros of $g_{22}$.

Before explaining  the content of the paper, 
we need to describe an algorithm for solving the global Liouville equation.
The subsequent algorithm 
is well-known in the context of the Liouville equation, and 
is based on the observation that if one  defines
 $L_\pm := \pa_\pm^2 Q/Q$,
then $\pa_\mp L_\pm =0$ on account of (1.1).
In terms of $\varphi$, 
$L_\pm$ are just the light cone components of the 
conformally improved Liouville stress-energy tensor.
For a smooth $Q$ which is periodic in $\sigma$,
$L_\pm$  are smooth, periodic 
functions  depending respectively on $x^\pm$ only.
For convenience,  the length of the period in $\sigma$ 
is set equal to $4\pi$.
As a first step for solving (1.1), one may therefore 
consider the pair of Schr\" odinger equations
$$
(\pa_\pm^2 -L_\pm )\psi^\pm =0
\eqno(1.3)$$
for some given smooth $L_\pm$ with $L_\pm (x^\pm +2\pi)= L_\pm (x^\pm)$.
This is also known as Hill's equation [\MW]. 
If $\psi^\pm_i(x^\pm)$ for $i=1,2$ are independent, real solutions 
normalized by the Wronskian condition
$$
\psi^\pm_2 \pa_\pm \psi^\pm_1 - \psi^\pm_1 \pa_\pm \psi^\pm_2=1,
\eqno(1.4)$$
then 
$$
Q (x^+, x^-):= \Psi_+ (x^+) \Psi_-^T (x^-), 
\qquad\hbox{where }\qquad 
\Psi_\pm := \pmatrix{ \psi_\pm^1&\psi_\pm^2},
\eqno(1.5)$$
solves (1.1).
This solution is automatically smooth as well, but its
periodicity in $\sigma$ constrains the admissible pairs $(L_+, L_-)$.
Namely, since $L_\pm$ are periodic, the solutions of the Hill's 
equations are quasiperiodic with some monodromy, i.e., 
$$
\Psi_\pm (x^\pm +2\pi) = \Psi_\pm (x^\pm) M_{\Psi_\pm} 
\qquad\hbox{for some}\qquad 
M_{\Psi_\pm} \in SL(2,\R).
\eqno(1.6)$$
The periodicity condition  $Q(x^+,x^-)=Q(x^++2\pi, x^--2\pi)$
is clearly equivalent to 
$$
M_{\Psi_-} = (M_{\Psi_+})^T.
\eqno(1.7)$$
Because the transformations
$$
\Psi_+ \mapsto \Psi_+ A,\quad
\Psi_- \mapsto \Psi_- (A^{-1})^T 
\qquad
A\in SL(2,\R  )
\eqno(1.8)$$
do  not change $Q$ and transform the monodromy matrices by conjugation,
one may assume without loss of generality that $M_{\Psi_+}$ 
 belongs to a given set of representatives of the conjugacy classes of 
$SL(2, \R)$ when writing the solutions of (1.1) in the
form (1.5).

The global Liouville equation is conformally invariant, since 
if $Q$ is a solution then so is $Q^{\alpha_+, \alpha_-}$ given by 
$$
Q^{\alpha_+,\alpha_-}(x^+,x^-):={1\over \sqrt{\pa_+\alpha_+(x^+)}}
{1\over \sqrt{\pa_- \alpha_-(x^-)}} Q(\alpha_+(x^+),\alpha_-(x^-))
\eqno(1.9)$$
for any conformal transformation 
$x^\pm \mapsto \alpha_\pm(x^\pm)$,
where the smooth functions $\alpha_\pm$ satisfy   
$$
\pa_\pm \alpha_\pm >0,
\qquad
\alpha_\pm (x^\pm + 2\pi) = \alpha_\pm(x^\pm) + 2\pi.
\eqno(1.10)$$
In order to describe all solutions of (1.1), it is enough to know 
the conformally nonequivalent ones.
An important fact is that the action of the 
conformal group on $Q$ in (1.9) is induced from its action on the
set of Hill's equations according to 
$$
L_\pm \mapsto L_\pm^{\alpha_\pm}:=
(\pa_\pm \alpha_\pm)^2 L_\pm\circ \alpha_\pm  -{1\over 2}
{\pa_\pm^3 \alpha_\pm \over \pa_\pm \alpha_\pm} +{3\over 4}
{(\pa_\pm^2 \alpha_\pm)^2 \over (\pa_\pm \alpha_\pm)^2}
\eqno(1.11{\rm a})$$
$$
\Psi_\pm \mapsto \Psi_\pm^{\alpha_\pm}:= 
{1\over\sqrt{\pa_\pm \alpha_\pm}}\Psi_\pm\circ \alpha_\pm.
\eqno(1.11{\rm b})$$
For each chirality, 
the action in (1.11a) is just the coadjoint action of the 
central  extension of the  conformal group ${\rm Diff}_0(S^1)$.
Its orbits in the space of `Virasoro densities' $L_\pm$ 
are known as the coadjoint 
orbits
of the Virasoro algebra\footnote{${}^{b}$}{\ninerm 
To be exact, the properly normalized 
Virasoro densities are $\L_\pm = 2\kappa L_\pm$.
Geometrically, they belong to  the hyperplane of 
centre  $C=24\pi \kappa$ in the dual of 
$\widehat {\rm diff}(S^1)$. See Section 2.}.
Equations (1.11a), (1.11b) together 
express the conformal covariance of  Hill's equation (1.3).

It is now clear that the classification of the Hill's equations 
under the conformal group is essentially the same as the
classification of the Virasoro coadjoint orbits,
and this is the crucial step for describing the solutions of
the global Liouville equation.
Indeed, since all smooth, periodic solutions of (1.1) arise from the
above algorithm, the conformally nonequivalent 
ones among these solutions of (1.1) can be written down explicitly 
once one has the list of conformally nonequivalent Virasoro densities, 
and the associated nonequivalent solutions of Hill's equation.

The classification of Virasoro coadjoint orbits is 
well-known [\LP,\Ki,\Seg,\Wi,\OK,\GRS,\G].
However, we found that the expositions that are available 
in  the literature are not explicit enough for what is needed 
in the study of the global Liouville equation.
This is especially true regarding representatives 
for the nonequivalent solutions of Hill's equation.
Therefore we have rederived the necessary classification
keeping in mind its application to study (1.1).
Our derivation is elementary, and we obtain complete proofs 
of all statements. 
Since our description of the classification 
might be useful in other applications too,
we thought it worthwhile to publish it as a review,
in which we included a comparison with the previous treatments of 
the problem.

The  above-mentioned review takes up the bigger part of
the paper, and then we come back to apply it to equation (1.1).
More precisely, the paper is organized as follows.
First, in Section 2, we summarize the WZNW description
of the global Liouville system.
The following section, which is the longest, is devoted 
to presenting  the list of nonequivalent Hill's equations 
together with their explicit solutions.
The reader is advised  to consult Subsection 3.7 for a summary.
Then, in Section 4, we describe the behaviour of the chiral 
energy functional (the zero mode of the Virasoro density) 
on the Virasoro coadjoint orbits.
This was also studied in Ref.~[\Wi], and we complete the
arguments of this reference on some minor points. 
In Section 5, we obtain an explicit 
list of solutions for the global Liouville equation,  
which  in particular shows 
the exact relationship between the topological type (number of zeros)
and the Virasoro orbit type of the solutions.
Combining this  with the results in Section 4,
we establish the behaviour of the `total energy'
(defined as the reduced WZNW Hamiltonian) 
in the topological sectors of equation (1.1).
Finally, we give our conclusions in Section 6, 
and there are  4 appendices containing  technical details.

\bigskip
\centerline 
{\bf 2.~Reduced WZNW treatment of the global Liouville equation}
\medskip

We here recall [\Bal,\FT] the interpretation 
of the global Liouville system as a reduced WZNW model.
This also serves to fix the conventions used in the subsequent 
sections.

Consider the WZNW model for the group $SL(2,\R)$ in the Hamiltonian 
framework. The phase space of the model (the set  of initial
data at $\tau=0$) can be  realized  as the manifold
$$
{\cal M}_{\rm WZ}=\{\, (g,J_+)\,\vert\, g\in C^\infty(S^1, SL(2,\R)),\,\,\,
J_+\in C^\infty(S^1, sl(2,\R))\,\}.
\eqno(2.1)$$
It comes equipped with the fundamental Poisson  brackets
$$\eqalign{
\{ g(\sigma), g(\bar \sigma)\}_{\rm WZ}=&0,\cr 
\{\tr(T_aJ_+)(\sigma), g(\bar \sigma)\}_{\rm  WZ}=& -T_a g(\sigma) 
\delta_{4\pi},\quad
\delta_{4\pi}=\delta_{4\pi}(\sigma -\bar \sigma),  \cr
\{\tr(T_a J_+)(\sigma), \tr(T_b J_+)(\bar \sigma)\}_{\rm WZ}=&
\tr([T_a, T_b] J)(\sigma) \delta_{4\pi}
 + 2\kappa \tr(T_a T_b) \delta_{4\pi}',\cr
}
\eqno(2.2)$$
where $\kappa\in \R$ is a nonzero constant, and  $T_a$ is 
a basis of $sl(2,\R)$.
The functions on $S^1$ are realized as periodic functions on $\R$ 
with period $4\pi$ using the variable $\sigma$ that  
parametrizes $S^1$ by $z=\exp({i \sigma \over 2})$,
and $\delta_{4\pi}(\sigma)={1\over 4\pi}\sum_{n\in {\bf Z}} 
e^{in\sigma/2} $ is the $4\pi$-periodic Dirac-$\delta$.
Defining  
$$
J_-= -g^{-1} J_+ g + 2\kappa  g^{-1} g',
\eqno(2.3)$$ 
the WZNW Hamiltonian 
$$
{\cal H}_{\rm WZ}= 
{1\over 4\kappa}\int_0^{4\pi} d\sigma\,\tr(J_+^2 +J_-^2)(\sigma)
\eqno(2.4)$$
generates the evolution equation 
$$
\kappa \pa_+ g = J_+ g,
\qquad 
\pa_- J_+ =0.
\eqno(2.5)$$
The global Liouville system will result from reduction
by the 
constraints\footnote{${}^{c}$}{\ninerm  
We could equivalently set the currents in (2.6) 
equal to any constants $\mu_\pm$ with $\mu_+ \mu_-<0$,
while $\mu_+\mu_- >0$ would lead to a system  
containing  the Liouville theory  with wrong sign.}
$$
\tr(e_{12} J_+) = 1, \quad
\tr(e_{21} J_-) =-1,
\eqno(2.6)$$
where the $e_{ij}$ are the standard matrices having $1$ for the 
$(i,j)$-entry and zero elsewhere.
These constraints are first class and generate the gauge transformations
$$
g \rightarrow \exp( A e_{12}) g \exp(-B e_{21}), 
\quad J_+\rightarrow 
\exp(A e_{12}) J_+ \exp(-A e_{12})  + 2\kappa A' e_{12},
\eqno(2.7)$$
for any  $A, B \in C^\infty(S^1, \R)$.
A globally well-defined gauge fixing is obtained by 
restricting the pair $(g,J_+)$ in such a way that   
$$
g= \pmatrix{R & P_+ \cr P_- & g_{22}\cr},
\quad
J_+=\pmatrix{ 0& \ell_+ \cr 1 & 0\cr},
\quad\hbox{and}\quad
J_-= -\pmatrix{0& 1 \cr \ell_- & 0\cr}.
\eqno(2.8)$$
Due to (2.3) these restrictions on $(g, J_+)$ can be rewritten in the form 
$$
R = \ell_+ g_{22} - 2\kappa P_+',
\quad
P_- = P_+ -2\kappa g_{22}',
\quad
\ell_- = R^2 - \ell_+ P_-^2 +2\kappa ( R' P_- - R  P_-'),
\eqno(2.9)$$
which express $R, P_-, \ell_-$ in terms of $g_{22}, P_+, \ell_+$, 
and a remaining constraint 
$$
1=\det g =g_{22}(\ell_+g_{22} - 2\kappa P_+') - P_+ 
(P_+ - 2\kappa  g_{22}').
\eqno(2.10)$$
Thus we can think of the reduced WZNW phase space 
as the submanifold in the space of 
the triples $(g_{22},P_+,\ell_+)$ of smooth,
periodic functions defined  by (2.10).

It is appropriate to comment here on the topological sectors 
of the reduced system.
Denote by $N[g_{22}]$ the number of zeros 
of $g_{22}(\sigma)$ over a period $0\leq \sigma < 4\pi$.
Note from (2.10) that $g_{22}(\sigma)$ and 
$g_{22}'(\sigma)$ cannot have simultaneous zeros: 
$$
\left(g_{22}^2 + (g_{22}')^2 \right)(\sigma) \neq 0.
\eqno(2.11)$$
Therefore $g_{22}(\sigma)$ cannot have any double zero, 
thus $N[g_{22}]$ is  
even (possibly $0$), and it defines a topological invariant of $g_{22}$.
Consequently, the reduced WZNW phase space 
decomposes into disconnected components labelled by the  
values of $N[g_{22}]$.
It is shown in Appendix A that the invariant $N[g_{22}]$ is inherited from 
the winding number $W[g]$ of $g: S^1 \rightarrow  SL(2,\R)$,
which is a topological invariant in the original WZNW model. 
The winding number is invariant with respect to the gauge
group in (2.7), since this group has trivial topology.

To describe the reduced WZNW dynamics,
we note that  the gauge fixing in (2.8) is preserved by the 
WZNW evolution equation (2.5), 
and therefore the reduced dynamics can be obtained 
by simply applying (2.5) to the gauge fixed variables $(g,J_+)$ in (2.8).
This yields the evolution equation
$$
\kappa \pa_+ g_{22} = P_+, \quad
\kappa \pa_+ P_+ = \ell_+ g_{22},
\quad
\pa_- \ell_+ =0.
\eqno(2.12)$$
In terms of the variables $(g_{22}, P_-, \ell_-)$,  
the reduced dynamics is equivalently given by 
$$
\kappa \pa_- g_{22} = P_-,
\quad
\kappa \pa_- P_- = \ell_- g_{22},
\quad
\pa_+ \ell_- =0.
\eqno(2.13)$$
Using (2.12-13) the constraint in (2.10) becomes 
$$
(Q \pa_+ \pa_- Q - \pa_+ Q \pa_- Q )=1 
\quad\hbox{where}\quad 
Q:=\kappa g_{22}.
\eqno(2.14)$$
This is just our global Liouville equation in (1.1).
Conversely, 
any  smooth, periodic solution  
of (2.14) gives a solution of (2.12) by   
defining  $P_+ := \pa_+ Q$ and $\ell_+ := \kappa^2 \pa_+^2 Q / Q$.
By this definition, (2.14) becomes just the constraint (2.10).
The smoothness of $\ell_+$ is also guaranteed since 
$\pa_+^2 Q / Q= \pa_+^2 \pa_- Q/ \pa_- Q$,
where $Q$ and $\pa_- Q$ cannot be simultaneously zero by (2.14).
In conclusion,  the global Liouville equation 
encodes the reduced WZNW dynamics.

We now wish to elaborate on the conformal symmetry of the 
reduced sytem. We first note that (2.12-13) imply 
$
\pa_\pm^2 Q(x^+, x^-) = L_\pm(x^\pm) Q(x^+,x^-)$
with 
$$ 
L_\pm(x^\pm)= {1\over \kappa^2} \ell_\pm(\sigma=\pm 2 x^\pm, \tau=0).
\eqno(2.15)$$
Hence, comparison with (1.3), (1.5) allows us to identify 
the variables $L_\pm$ 
in (2.15) with the similarly named variables  
of Section 1.
As $\ell_\pm$ are  functions on the reduced WZNW phase space that
carries a natural induced Poisson bracket (the Dirac bracket), 
we can then compute the Poisson brackets of 
$$
\L_\pm(x^\pm) := 2\kappa L_\pm(x^\pm),
\eqno(2.16)$$
and find [\Bal] 
$$
\{ \L_\pm(x), \L_\pm(y)\}= - \kappa 
\delta_{2\pi}'''(x-y) +
2\L_\pm(y) \delta_{2\pi}'(x-y)
- \L_\pm'(y) \delta_{2\pi}(x-y),
\eqno(2.17)$$
where $\delta_{2\pi}(x)={1\over 2\pi} \sum_{n\in {\bf Z}} e^{inx}$
is the $2\pi$-periodic Dirac-$\delta$.
Defining  the modes $\L_\pm^n$ by
$$
\L_\pm^n:= \int_0^{2\pi} dx\, e^{inx} \L_\pm(x)+ \kappa \pi \delta_{n,0},
\qquad \forall n\in {\bf Z},
\eqno(2.18)$$
eq.~(2.17) is equivalent to 
$$
\{ \L_\pm^n, \L_\pm^m\} =(-i) \left( (n-m) \L_\pm^{n+m} + 24\pi  \kappa 
{n(n^2 -1) \over 12} \delta_{n+m, 0}\right).
\eqno(2.19)$$
These are two commuting copies of the Virasoro algebra with centre 
$$
C= 24\pi \kappa.
\eqno(2.20)$$
The overall factor $(-i)$ on the r.h.s.\ of (2.19) reflects correctly 
the correspondence principle between Poisson brackets and commutators,
and the second term has the standard form that vanishes for 
the M\"obius subalgebras generated by $\L^{-1}_\pm, \L_\pm^0, \L_\pm^1$ 
due to the shift in the definition of $\L_\pm^0$ in (2.18). 

The variation $\Delta_{\epsilon_\pm}$ of a dynamical variable 
under an infinitesimal conformal transformation
$x^\pm \mapsto x^\pm + \epsilon_\pm(x^\pm)$
is defined by its Poisson bracket with the respective charge 
$T_{\epsilon_\pm}=\int_0^{2 \pi} dx \epsilon_\pm(x) \L_\pm(x)$.
In particular, this gives 
$$
\Delta_{\epsilon_{\pm}} L_\pm := -\{ T_{\epsilon_\pm}, L_\pm \} 
= \epsilon_\pm L'_\pm + 2\epsilon_\pm' L_\pm -
{1\over 2}\epsilon_\pm''',
\eqno(2.21)$$
which we recognize to be the infinitesimal version of the
transformation rule in (1.11a).
It follows that 
as the variable $L_\pm$ maps out 
an orbit of the conformal group according to (1.11a), 
the  variable $\L_\pm$ in (2.16) 
runs through a coadjoint orbit of the Virasoro algebra
at centre $C=24\pi \kappa$. 
The advantage of using the variables $L_\pm$ is that 
their transformation rule is independent of $C$, and this 
allows one to describe the Virasoro coadjoint
orbits at any $C\neq 0$ at one stroke.
We shall use this device  in the subsequent sections,
but for proper interpretation the relationship (2.16)
must be remembered.

Let us further observe that the WZNW Hamiltonian (2.4) survives the
reduction because its restriction to the constrained manifold  
(2.6) is gauge invariant. 
This gauge invariant function  defines 
the natural Hamiltonian for the reduced WZNW model, that is 
for the global Liouville system. 
The  Hamiltonian is of course constant along any solution 
of the reduced dynamics.
Denoting its value by  ${\cal H}_{\rm WZ}[Q]$ we easily verify that 
$$
{\cal H}_{\rm WZ}[Q] = {1\over 2} \left(\L_+^0 + \L_-^0\right) -\kappa \pi,
\eqno(2.22)$$
where $\L_\pm^0$ are  defined in (2.18).
On account of (2.16), 
it is convenient 
to represent the Virasoro zero modes in the form 
$$
\L^0_\pm = 4\pi\kappa \left( E[L_\pm] + {1\over 4}\right)
\quad\hbox{where}\quad
E[L_\pm]=
{1\over 2\pi} \int_0^{2\pi} dx L_\pm(x). 
\eqno(2.23)$$
Later we will study the question of boundedness from below 
for the  `chiral energy functionals' $E[L_\pm]$ on the Virasoro 
coadjoint orbits, and apply the result to understand the behaviour
of the `total energy' ${\cal H}_{\rm WZ}[Q]$ in the global Liouville system.

\bigskip 
\centerline{\bf 3.\  The normal forms of Hill's equation}
\medskip

In this section, we obtain the 
classification of the Virasoro coadjoint orbits 
by analysing the Hill's equation associated with the Virasoro densities.
We first define a rough classification 
by attaching to each Virasoro orbit the conjugacy class of the monodromy 
matrix of the corresponding Hill's equation. 
We then find  the  conformally nonequivalent Virasoro 
densities, and the solutions of Hill's equation  with any given monodromy. 
The main ideas and the results of this classification are known,
but our presentation is different,  since we list 
explicit representatives for the nonequivalent objects.

\vfill\eject

\medskip
\noindent 
{\bf 3.1.\ Statement of the problem and recall of some known results}
\medskip

We are interested in the classification of the Hill's equations 
$$
\psi''=L\psi,
\eqno(3.1)$$
where $L$ is a smooth, real, periodic function on the line $\R$  with 
period $2\pi$, with respect to the group of transformations
$$
L\mapsto 
L^\alpha :={\alpha'}^2 L\circ \alpha +S(\alpha),
\qquad
S(\alpha):=-{1\over 2}{\alpha'''\over \alpha'}+{3\over 4}
{{\alpha''}^2\over {\alpha'}^2},
\eqno(3.2{\rm a})$$
$$
\psi\mapsto 
\psi^\alpha := {1\over \sqrt{\alpha'}}\psi\circ\alpha,
\eqno(3.2{\rm b})$$
where $\alpha$ is a smooth, real function
on ${\bf R}$ with the property
$$
\alpha'>0,
\quad
\alpha(x+2\pi)=\alpha(x)+2\pi.
\eqno(3.3)$$
The mappings $\alpha:{\R}\rightarrow  {\R}$ with this
property form a subgroup of the group
${\rm Diff}_0({\R})$ of orientation 
preserving diffeomorphisms of the line.
This group is denoted by 
${\widetilde {\rm Diff}}_0(S^1)$
since it is a covering group of the group
of orientation preserving diffeomorphisms of
the circle ${\rm Diff}_0(S^1)$. 
The homomorphism 
$$
\chi: {\widetilde {\rm Diff}}_0(S^1)\rightarrow {\rm Diff}_0(S^1)
\eqno(3.4)$$
is naturally induced from the map
$
{\R}\rightarrow  S^1$
given by 
$x\mapsto e^{ix}$.
The kernel of $\chi$ is the group of translations on  ${\R}$ by 
integer multiples of $2\pi$.
Formula (3.2) defines an action of
${\widetilde {\rm Diff}}_0 (S^1)$ on the set of Hill's equations, and  
we wish to find a list of  `normal forms' of $L$ and those
of the corresponding solutions $\psi$ that are 
nonequivalent under this group.

The action of $\widetilde{\rm Diff}_0(S^1)$ in (3.2{\rm a}) 
on the space of $L$'s factors 
through  the homomorphism $\chi$ to an action of ${\rm Diff}_0(S^1)$.
By identifying the space of $L$'s with a hyperplane 
at centre $C\neq 0$ in the dual of the Virasoro Lie algebra,
this action can be recognized as the coadjoint action of
the central extension of ${\rm Diff}_0(S^1)$, 
which at the infinitesimal level is the coadjoint action of the 
Virasoro Lie algebra (see e.g.~[\Wi]).
Thus our problem to classify the orbits of the action in (3.2) is 
essentially equivalent to the classification  of the Virasoro coadjoint 
orbits [\LP,\Ki,\Seg,\Wi,\OK,\GRS,\G].
We shall present the solution in an
approach convenient  for application to equation (1.1).

We wish to stress here that in this paper we often 
call the orbits of the group action in 
(3.2a) Virasoro coadjoint orbits,
even though it is $\L={C\over 12 \pi}L$ which is the
properly normalized Virasoro density 
(see eqs.~(2.16)--(2.21)).
In this way,  we effectively describe the coadjoint 
orbits at an arbitrary $C\neq 0$ in  $C$--independent terms,
and this slight abuse of notation should 
not lead to any misunderstanding.

For purposes of comparison between our approach, which is close 
to Refs.~[\LP,\Seg,\OK],  and the one described by Kirillov [\Ki]
and Witten [\Wi],  we now recall some known results.
First note that the Lie algebra of the little group  
$G[L]\subset {\rm Diff}_0(S^1)$ of $L$ is spanned 
by the $2\pi$-periodic vector fields
 $\xi(x) {\partial \over \partial x}$ (infinitesimal conformal
transformations) that solve 
$$
\xi''' -4L \xi' -2L' \xi =0,
\eqno(3.5)$$
and if $\psi_1$, $\psi_2$ are the independent solutions of (3.1),
then the  solutions of (3.5) are the products
$$
\psi_1^2,
\quad
\psi_2^2,
\quad
\psi_1\psi_2,
\eqno(3.6)$$
which admit either 1 or 3 periodic linear combinations [\MW,\Ki].
The coadjoint orbit ${\cal O}_L$ through  $L$ can be represented as
$$
{\cal O}_L ={\rm Diff}_0(S^1)/G[L]=
{\widetilde {\rm Diff}}_0(S^1)/\tilde G[L],
\eqno(3.7)$$
where 
$$
\tilde G[L]:=\chi^{-1}\left(G[L]\right)\subset 
{\widetilde {\rm Diff}}_0(S^1)
\eqno(3.8)$$
is the lift of $G[L]\subset {\rm Diff}_0(S^1)$. 
In the terminology of Ref.~[\Wi], 
the nonconjugate little groups 
$G[L]\subset {\rm Diff}_0(S^1)$ that define the possible orbits 
${\cal O}_L$  are [\Ki,\Wi] 
$$
S^1,
\quad
PSL^{(n)}(2,\R),
\quad
T_{n,\Delta},
\quad
{\tilde T}_{n,\pm}  \qquad (n\in {\bf N},\quad \Delta >0). 
\eqno(3.9)$$
Here $S^1$ is the group of rigid rotations of
$S^1$; $PSL^{(n)}(2,\R)$ is a subgroup of 
${\rm Diff_0}(S^1)$ isomorphic to the $n$-fold cover 
of the projective group $PSL(2,\R)=SL(2,\R)/Z_2$; 
${T}_{n,\Delta}$ and  ${\tilde T}_{n,\pm}$  are special 
one dimensional subgroups whose Lie algebras are given
by vector fields on $S^1$ having $2n$ simple and, 
respectively, $n$ double zeros.
The groups 
$T_{n,\Delta}$ and ${\tilde T}_{n,\pm}$
are not connected, since they also contain  certain discrete 
subgroups isomorphic to the group 
${\bf Z}_n$  of rigid rotations by multiples of the 
angle $2\pi/n$.
These little groups  classify  the coadjoint orbits
up to diffeomorphisms, but conjugate
little groups may belong to nonequivalent 
points in the space of $L$'s.
This  actually happens only for the
orbit ${\rm Diff}_0(S^1)/S^1$, which carries a
one parameter family of nonequivalent symplectic structures
at fixed value of the central parameter.

The subgroup $\tilde G[L]$ in (3.8) still acts on the solutions of 
Hill's equation at the standard point $L$ of ${\cal O}_L$, and 
this has to be taken into account when listing the conformally
nonequivalent solutions.
We shall present a complete list of conformally nonequivalent solutions 
of Hill's equation under the additional assumptions that the
solutions are normalized by a Wronskian condition and that they have
nonconjugate monodromy matrices.
The monodromy matrix will be fixed to vary in  
a set of representatives of the conjugacy classes of $SL(2,\R)$ chosen 
in the next subsection.

\medskip
\noindent{\bf 3.2.\  Rough classification by 
conjugacy classes of the monodromy matrix}
\medskip

For given $L$, let $\psi_i$ ($i=1,2$) be independent,
real solutions of  Hill's equation 
normalized by the Wronskian condition
$$
\psi_2 \psi_1'-\psi_2'\psi_1=1.
\eqno(3.10)$$
The solution space of (3.1) is two-dimensional and
the {\it solution vector}\footnote{${}^{d}$}{\ninerm Solution {\it basis} 
would be  a more correct term, but we often operate on $\Psi$ 
 like on a 2-vector.} 
$\Psi$ given by
$$
\Psi:=\pmatrix{\psi_1 &\psi_2 }
\eqno(3.11)$$
is determined up to
$$
\Psi\mapsto \Psi A,
\qquad 
\forall \,A\in SL(2,{\bf R}). 
\eqno(3.12)$$ 
Since $L$ is $2\pi$-periodic, the translation operator
$$
\Psi \mapsto \tilde \Psi,
\quad
\tilde\Psi(x):=\Psi(x+2\pi)
\eqno(3.13)$$
acts on the solutions, and it obviously preserves
the Wronskian condition (3.10). 
Thus we can associate the {\it monodromy matrix}
$M_\Psi\in SL(2,\R)$ to $\Psi$ by
$$
\tilde \Psi = \Psi M_\Psi.
\eqno(3.14)$$
The monodromy matrix enjoys the property
$$
M_{\Psi A}=A^{-1} M_\Psi A,
\quad 
\forall A\in SL(2,{\bf R}).
\eqno(3.15)$$   
This means that the conjugacy class 
of the monodromy matrix depends only on 
the potential $L$ defining the Hill's equation
(3.1), and is independent of the freedom in (3.12). 
Moreover, 
if $\Psi$ 
is a solution vector of the Hill's equation
at $L$, then
$$
\Psi^\alpha := \pmatrix{\psi_1^\alpha &\psi_2^\alpha}
\eqno(3.16)$$
is a solution vector at $L^\alpha$
(it satisfies the Wronskian condition) with 
$$
M_{\Psi^\alpha}=M_{\Psi}.
\eqno(3.17)$$
In conclusion, the {\it conjugacy class} of the monodromy
matrix is a {\it conformally invariant} function of $L$.

Considering the action of the little group 
$\tilde G[L]$ in (3.8),  (3.17) implies that  
$$
\Psi^{\alpha^{-1}} = \Psi \gamma(\alpha) \qquad\quad  
\forall\, \alpha\in \tilde G[L],
\eqno(3.18{\rm a})$$
where $\gamma(\alpha)$ belongs to the little group 
$G[M_\Psi]\subset SL(2,\R)$ of the monodromy matrix.
For any fixed $\Psi$, the map 
$$
\gamma: \tilde G[L]\rightarrow  G[M_\Psi] 
\eqno(3.18{\rm b})$$
is actually a {\it homomorphism} (this is why we used the 
inverse in the definition (3.18a)).
The normalized 
solution vectors of Hill's equation at the standard point 
$L\in {\cal O}_L$, with fixed monodromy matrix $M_\Psi$
are  in one-to-one correspondence with the elements $g$ of
$G[M_\Psi]$,  by $g\leftrightarrow \Psi g$.
The orbits of the little group 
$\tilde G[L]$ in the space of these solution vectors
can therefore be parametrized by the points of the coset space
$$
\gamma(\tilde G[L])\backslash G[M_\Psi].  
\eqno(3.19)$$
We shall see that this space consists of 
a single point in most cases, but at most 2 points.

Now let ${\cal M}$ be the set of conjugacy classes
of $SL(2,{\R})$ and
$$
p: SL(2,{\R})\rightarrow  {\cal M}
\eqno(3.20)$$
be the canonical projection.
Denote by  ${\cal V}$ the set
of the Virasoro coadjoint orbits ${\cal O}_L$.
We have a well-defined map from
${\cal V}$ to ${\cal M}$ 
given by the formula
$$
{\cal O}_L\mapsto p(M_{\Psi}).
\eqno(3.21)$$
This map is many-to-one, and yields a
rough classification of the Virasoro coadjoint orbits.
To describe it more concretely,  we  below present  
a list of representatives
for  the conjugacy classes of $SL(2,{\R})$.
There exist 4 types of conjugacy classes: elliptic, hyperbolic, 
parabolic and one-point classes. 
The  elliptic and hyperbolic classes are generic since they contain the 
elements $g\in SL(2,\R)$ for which $\vert {\rm tr\,}(g)\vert <2$ 
and $\vert {\rm tr\,}(g)\vert >2$, respectively.
Our terminology is somewhat 
nonstandard concerning the remaining special classes,
as we call parabolic only those that contain
more than one element. 

Our representatives are as follows.
First, a hyperbolic conjugacy class  
 is represented
by a matrix $B_\eta(b)$ of the form 
$$
B_\eta(b):=\eta \pmatrix{e^{2\pi b}&0\cr 0& e^{-2\pi b}\cr},
\qquad \eta=\pm 1, \quad b>0.
\eqno(3.22{\rm a})$$
The corresponding little group $G[B_\eta(b)]\subset SL(2,{\R})$ 
contains the matrices 
$$
\pmatrix{\beta&0\cr 0&{1/\beta}\cr},
\qquad
\beta\neq 0.
\eqno(3.22{\rm b})$$
Second,  an elliptic  class  
is represented by a  matrix $C(\omega)$ of the form
$$
C(\omega):=\pmatrix{\cos\omega&-\sin\omega\cr\sin\omega&\cos\omega},
\qquad
0<\omega< \pi,
\quad
\pi<\omega<2\pi.
\eqno(3.23{\rm a})$$
The little group $G[{C(\omega)}]$  
contains  the matrices  
$$
\pmatrix{\cos\vartheta&-\sin\vartheta\cr\sin\vartheta&\cos\vartheta},
\qquad
0\leq\vartheta<2\pi.
\eqno(3.23{\rm b})$$
Third, one has the $2$  one-point classes given  by 
$$
E_\eta:=\eta \pmatrix{1&0\cr 0& 1\cr},
\qquad
\eta=\pm 1,
\eqno(3.24)$$
whose little group is the full group 
$G[E_\eta]=SL(2,{\R})$.
Finally, one has the $4$  parabolic  classes represented 
by the matrices $P^q_\eta$,
$$
P^q_\eta:=\eta\pmatrix{1&0\cr q&1\cr},
\qquad \eta=\pm 1,\quad q=\pm 1. 
\eqno(3.25{\rm a})$$
The little group $G[P^q_\eta]$ is the maximal nilpotent 
subgroup consisting of the matrices 
$$
\pmatrix{\pm 1&0\cr \gamma&\pm1\cr},
\qquad
\forall \gamma\in {\bf R}.
\eqno(3.25{\rm b})$$

In  addition to the monodromy parameter
furnished by the map (3.21), the classification
of the Virasoro coadjoint orbits requires an
additional discrete parameter [\Seg,\OK]. 
The orbits associated with  monodromy conjugacy classes of type
$B$, $C$, $E$, $P$ will be denoted
by the letters ${\cal B}$, ${\cal C}$,
${\cal E}$, ${\cal P}$, respectively, 
plus parameters needed 
for their complete specification.
This will be related to 
the notation in Refs.~[\LP,\Ki,\Wi].  

\medskip
\noindent
{\bf 3.3.\ Virasoro coadjoint orbits with elliptic monodromy}
\medskip

If $L$ is such that the monodromy of Hill's equation
is elliptic,  choose a solution vector 
$\Psi=\pmatrix{\psi_1&\psi_2}$ 
so that the monodromy matrix is of the form $C(\omega)$ in (3.23a).
Then 
$$
\cR:=\psi_1^2 + \psi_2^2>0 
\eqno(3.26)$$
is a {\it periodic} function,
which is a well-defined functional of $L$,  
since it is unchanged under $\Psi\mapsto \Psi A_0$,
$A_0\in G[C(\omega)]$.
It is straightforward to verify the equality
$$
L={\psi_1''\psi_1 +\psi_2''\psi_2\over \psi_1^2 +\psi_2^2}
= {1\over 2}{\cR''\over \cR}-{1\over 4}{{\cR'}^2 \over 
\cR^2}-{1\over \cR^2}.
\eqno(3.27)$$
It is also clear that  
$$
\nu:={1\over \pi}\int_0^{2\pi} {dy\over \cR(y)}\,,
\qquad
\nu >0,
\eqno(3.28)$$
is invariant  under the conformal group (3.2),
and that 
$$
\alpha(x):={2 \over \nu}\int_0^x\,{dy\over \cR(y)}
\eqno(3.29)$$
satisfies (3.3).
With this $\alpha$, (3.27)
can be rewritten as the equality
$$
L={\alpha'}^2 \left({-{\nu^2\over 4}}\right) + S(\alpha).
\eqno(3.30)$$
By (3.2),  this means that $L$ lies
on the Virasoro coadjoint orbit through 
$$
L_\nu:=-{\nu^2\over 4},
\eqno(3.31)$$
i.e., $L=L_\nu^\alpha$.
Hence the `standard Virasoro
densities' $L_\nu$ provide a complete,
nonredundant set  of {\it representatives}
of the Virasoro coadjoint orbits with 
elliptic monodromy.

The solution vector
$\Psi_\nu=\pmatrix{\psi_{1,\nu}& \psi_{2,\nu}}$ 
of Hill's equation at $L_\nu$, 
with standard monodromy, is given by
$$
\psi_{1,\nu}(x)={\sqrt{2\over \nu}\sin{\nu x\over 2}},
\qquad
\psi_{2,\nu}(x)={\sqrt{2\over \nu}\cos{\nu x\over 2}},
\eqno(3.32)$$
up to the freedom contained in the little group $G[C(\omega)]$.
Indeed, the corresponding monodromy matrix is $C(\omega)$ with
$$
\omega=\nu\pi -2\pi d
\quad \hbox{ for some }\quad d\in {\bf Z}_+:=\{ 0\}\cup {\bf N}.
\eqno(3.33)$$
This establishes the relationship between
$\nu$ and the monodromy parameter $\omega$,
and one sees that the latter determines the
former up to a nonnegtive integer $d$.
The integer just mentioned may be regarded as 
the discrete parameter that labels the orbits with elliptic monodromy
in addition to $\omega$.
Note that $\nu\notin {\bf N}$ since 
we have elliptic monodromy (3.23a).

The little group $G[L_\nu]\subset {\rm Diff}_0(S^1)$
is in fact [\Ki,\Wi] the group of rigid rotations of $S^1$, which
lifts to the translation group of ${\bf R}$, i.e.,
${\tilde G}[L_\nu]=\R$ (see also Appendix B).
One can easily  evaluate the action of this group on the
solution vector $\Psi_\nu$ in (3.32),
which shows that the orbit of $\tilde G[L_\nu]$ contains  all
solution vectors at $L_\nu$ with the same monodromy matrix,
i.e.,  the coset space in (3.19) consists of a single point in this case.
We conclude that the solution in (3.32) is a 
representative for all solutions
of Hill's equation with monodromy $C(\omega)$ and fixed value 
of the invariant $\nu$ in (3.28).
In particular, since $L=L_\nu^\alpha$, 
we can  write
$$
\psi_1=\psi_{1,\nu}^\alpha =
 \sqrt{{2 \over \nu \alpha'}}\sin {\nu \over 2}\alpha,
\quad
\psi_2=\psi_{2,\nu}^\alpha =
\sqrt{{2\over \nu \alpha'}}\cos {\nu \over 2}\alpha,
\eqno(3.34)$$
up to the little group of conformal transformations.

To summarize, 
the Virasoro coadjoint orbits with
elliptic monodromy are
$$
{\cal C}(\nu):={\cal O}_{L_\nu}
={\rm Diff_0}(S^1)/S^1={\widetilde{\rm Diff}}_0(S^1)/{\bf R},
\qquad
 \nu>0,
\quad \nu \notin {\bf N},
\eqno(3.35)$$
and, for any $L\in {\cal C}(\nu)$,
the Wronskian-normalized solutions of Hill's equation with monodromy
matrix $C(\omega)$  are  conformally equivalent 
to the solution in (3.32).
The orbits in (3.35) associated with different values of $\nu$ 
are diffeomorphic, but of course
this does not mean equivalence from the
phase space (symplectic form, energy etc) point of view.

\medskip
\noindent 
{\bf 3.4.\  Virasoro coadjoint orbits with 
$E_\eta$-type monodromy}
\medskip

The analogues of equations (3.26-32) are valid in this case as well.
We find that any $L$ of
monodromy type $E_\eta$ in (3.24) lies on a Virasoro orbit through
$L_\nu =-{\nu^2\over 4}$ for some positive $\nu$.
The monodromy matrix of the normalized solution vector (3.32) at $L_\nu$ 
becomes the matrix $E_\eta$ if $\nu$ is an integer.
Therefore we have 
$$
\nu=n \in {\bf N}
\quad \hbox{and}\quad
\eta=(-1)^n.
\eqno(3.36)$$
A difference from the elliptic case is that $\cR$ in (3.26) is 
{\it not unique} in the present case, since now 
the solution vector $\Psi$ can be transformed according to  
$$
\Psi\mapsto \Psi A,
\qquad
\forall\, A\in G[E_\eta]=SL(2,{\R}),
\eqno(3.37)$$
and $\cR$ is not invariant under such a transformation in general.

The Lie algebra of the little group $G[L_n]\subset {\rm Diff}_0(S^1)$ is 
spanned by the vector fields 
$$
{\partial \over \partial x},
\quad 
\cos nx {\partial \over \partial x},
\quad
\sin nx {\partial \over \partial x},
\eqno(3.38)$$
which define nonconjugate embeddings of $sl(2,\R)$ into the Lie algebra of
${\rm Diff_0}(S^1)$ for different values of  $n$.
Globally [\Ki,\Wi] (see also Appendix B),  
$G[L_n]$ is the $n$-fold covering group  of $PSL(2,{\R})$, 
denoted by $PSL^{(n)}(2,{\bf R})$.
Then equations (3.8), (3.18) yield a homomorphism 
$$
\gamma: \chi^{-1}\bigl( PSL^{(n)}(2,{\R})\bigr)\rightarrow  SL(2,{\bf R}),
\eqno(3.39)$$
which  is  {\it surjective}, since it is just the 
standard homomorphism of the universal covering group of $SL(2,\R)$ 
onto $SL(2,\R)$.
A particular consequence is that, despite the noninvariance of $\cR$ 
under (3.37), 
$$
n={1\over \pi} \int_0^{2\pi} {dy\over \psi_1^2 +\psi_2^2}
\eqno(3.40)$$
is independent from the choice of the solution vector $\Psi$,
as it should be.
This integral is obviously invariant under conformal
transformations, and  this implies its invariance under (3.37), since 
the latter transformations arise from conformal transformations 
by the surjective homomorphism (3.39).
Another consequence is that the coset space (3.19) consists 
of a single point for $M_\Psi=E_\eta$.

In conclusion, the Virasoro coadjoint orbits with 
monodromy matrix $E_\eta$ are 
$$
{\cal E}_n:={\cal O}_{L_n}={\rm Diff}_0(S^1)/ PSL^{(n)}(2,{\R})=
{\widetilde{\rm Diff}}_0(S^1)/\chi^{-1}(PSL^{(n)}(2,{\bf R}))
\eqno(3.41)$$
where $n\in {\bf N}$ and $\eta=(-1)^n$. 
The Wronskian-normalized solutions  of Hill's equation (3.1) at  
$L\in {\cal E}_n$ are conformally equivalent to the  solution 
$\Psi_n$ at $L_n$ given by  
$$
\psi_{1,n}(x)={\sqrt{2\over n}\sin{n x\over 2}},
\qquad
\psi_{2,n}(x)={\sqrt{2\over n}\cos{n x\over 2}}.
\eqno(3.42)$$

\medskip
\noindent 
{\bf 3.5.\ Virasoro coadjoint orbits with hyperbolic monodromy}
\medskip

If $L$ is such that the monodromy of Hill's equation is hyperbolic,
consider a solution vector $\Psi=\pmatrix{\psi_1 &\psi_2}$ whose
monodromy matrix takes the standard form $B_\eta(b)$ in (3.22a).
Then associate a nonnegative integer $n$ to $L$ by 
means of\footnote{${}^{e}$}{\ninerm Because of the Wronskian condition,
one could equivalently use $\psi_1$ in the definition.}
$$
L\mapsto n(L):=
\hbox{number of zeros of $\psi_2(x)$ for $0\leq x <2\pi$}.
\eqno(3.43)$$
Since $\Psi$ is unique up to 
$$
\Psi \mapsto \Psi A,
\qquad
A\in G[B_\eta(b)],
\eqno(3.44)$$
the map  $L\mapsto n(L)$ is well-defined.
Furthermore, it follows from the transformation rule (3.2) 
that $n(L)$ is a conformally {\it invariant} function 
on the space of $L$'s with hyperbolic monodromy.
The discrete invariant 
$n\in {\bf Z}_+$ 
determines the invariant $\eta$ appearing in the
specification of the monodromy matrix, 
$$
\eta = (-1)^n.
\eqno(3.45)$$
Below, we show that the monodromy invariant $b>0$ and
$n$ together provide a complete classification 
of the Virasoro coadjoint orbits with hyperbolic monodromy,
which may therefore be labelled as
$$
{\cal B}_n(b),
\qquad
b>0,\quad n\in {\bf Z}_+.
\eqno(3.46)$$
We proceed by proving that any $L$ with
invariants $b$, $n$ can be brought  to a
standard form by a conformal transformation.
We first deal with the simplest case.

\smallskip
\noindent 
{\it 3.5.1.\ The case ${\cal B}_0(b)$.}
Since $\psi_2$ has no zeros, the function $u$ defined by
$$
u:={\psi_1 \over \psi_2}
\eqno(3.47)$$
is smooth in this case.
This function satisfies
$$
u'={1\over \psi_2^2}>0
\quad\hbox{and}\quad
u(x+2\pi)=e^{4\pi b} u(x),
\eqno(3.48)$$
and hence it is monotonically increasing with limits 
$$
\lim_{x\to - \infty} u(x) =0,
\quad  
\lim_{x\to +\infty}u(x)=+\infty.
\eqno(3.49)$$
These properties of $u$  permit to define an element 
$\alpha \in {\widetilde {\rm Diff}}_0(S^1)$
by 
$$
\alpha := {1\over 2b} \ln u.
\eqno(3.50)$$
It is  then straightforward to verify the identity 
$$
L={\psi_2''\over \psi_2}=b^2 (\alpha')^2 +S(\alpha),
\eqno(3.51)$$
proving  that $L$ lies on the orbit 
of the `standard point' 
$$
L_{2ib} := b^2.
\eqno(3.52)$$
The little group  $G[L_{2ib}]$ is the group of rigid 
rotations of $S^1$.
Therefore we have 
$$
{\cal B}_0(b)={\cal O}_{L_{2ib}}={\rm Diff}_0(S^1)/S^1
={\widetilde {\rm Diff}}_0(S^1)/{\bf R},
\eqno(3.53)$$
where we used that 
$$
\tilde G[L_{2ib}]={\bf R}
\eqno(3.54)$$
is the translation group of ${\bf R}$.
We further claim that the {\it conformally nonequivalent}
solutions of Hill's equation at $L_{2ib}$ are given by 
$\Psi^{\pm }=\pmatrix{\psi_1^\pm &\psi_2^\pm}$, where 
$$
\psi_1^\pm(x)=\pm {1\over \sqrt{2b}} e^{bx},
\quad
\psi_2^\pm (x)=\pm {1\over \sqrt{2b}} e^{-bx}.
\eqno(3.55)$$
Indeed, the image of the little group (3.54) under the homomorphism
(3.18) is the connected component of
the identity in $G[B_\eta(b)]$, and thus   
the coset space (3.19) now consists of 2 points.
As a consequence, the 
solutions of Hill's
equation with  monodromy matrix $B_+(b)$ at any $L\in {\cal B}_0(b)$ 
are conform transforms of $\Psi^\pm$. 
  
\smallskip
\noindent
{\it 3.5.2.\ The case ${\cal B}_n(b)$ for $n\in {\bf N}$.} 
We start by noting that
if $\Psi=\pmatrix{\psi_1 &\psi_2}$ is a pair of smooth 
functions subject to the Wronskian condition (3.10), then
$$
L:={\psi_1''\over \psi_1}={\psi_2''\over \psi_2}
\eqno(3.56)$$ 
is also smooth.
If in addition 
$$
\Psi(x+2\pi)=\Psi(x) B_\eta(b),
\eqno(3.57)$$ 
then $L$ is periodic as well. 
In other words, $\Psi$ is a solution vector of Hill's equation at $L$.

To show that Virasoro coadjoint orbits with $n(L)\neq 0$  exist, 
it is enough to present examples.
For  arbitrarily chosen $b>0$ and $n\in {\bf N}$, let us define
$$\eqalign{
\psi_1(x):&={e^{bx}\over \sqrt{F(x)}}\sqrt{n\over 2}
\left( {2b\over n^2}\cos {nx\over 2} +{2\over n} 
\sin {nx\over 2}\right) \cr
\psi_2(x):&={e^{-bx}\over \sqrt{F(x)}}\sqrt{2\over n} \cos {nx\over 2}\cr}
\eqno(3.58{\rm a})$$
where
$$
F(x)=F_{n,b}(x):= \cos^2{nx\over 2} +
\left( \sin {nx\over 2} +{2b\over n}\cos {nx\over 2}\right)^2.
\eqno(3.58{\rm b})$$
Since the function $F$ has no zeros, $\Psi$ is smooth. 
It is clear that $\psi_2$ has $n$ zeros, and (3.10), (3.57) are satisfied.
Thus $\Psi$ is a solution vector of Hill's equation at $L_{n,b}:=L$  
given by (3.56), which explicitly reads as 
$$
L_{n,b}=b^2 +{n^2 + 4b^2 \over 2F} -
{3\over 4}{n^2 \over F^2}.
\eqno(3.59)$$
This Virasoro density and the solution $\Psi$ in (3.58) 
are deformations of the ${\cal E}_n$-type 
representatives $L_n=-{n^2\over 4}$ and $\Psi_n$ in (3.42)
that are recovered in the limit $b=0$.

We now claim that  any Virasoro density with hyperbolic monodromy
and discrete invariant  $n\in {\bf N}$ lies on the
orbit ${\cal B}_n(b)$ of the `standard point' $L_{n,b}$ in (3.59).
The idea of the proof is as follows. 
For an arbitrary $\bar L$  with the same monodromy
and discrete invariants as $L_{n,b}$,
let $\bar \Psi=\pmatrix{\bar \psi_1 & \bar \psi_2}$ be
a  normalized solution vector at $\bar L$ with monodromy matrix $B_\eta(b)$.
Using that both $\psi_2$ in (3.58) and $\bar \psi_2$ have $n$ zeros,
it is not hard to see that $\bar \psi_2$ 
can be transformed into $\psi_2$ by a conformal transformation, i.e.,
$$
\exists\, \alpha\in {\widetilde{\rm Diff}}_0(S^1)
\quad \hbox{such that} \quad 
{1\over \sqrt{\alpha'}} {\bar \psi}_2(\alpha(x))=\psi_2(x).
\eqno(3.60)$$
A construction of the required $\alpha$ is contained in Appendix C.
Once (3.60) is proven, 
$L=\bar L^\alpha$ follows from Hill's equation.
Since $\bar \psi_2$ uniquely determines $\bar \psi_1$
by the Wronskian and monodromy conditions,
this argument  also proves  that the solution 
$\bar \Psi$ is conformally equivalent to the solution $\Psi$ in (3.58).

The little group $G[L_{n,b}]\subset {\rm Diff}_0(S^1)$ is spanned
by the vector field $\xi(x){\partial \over \partial x}$, where 
$$
\xi(x)=\psi_1(x)\psi_2(x)=\cos {nx\over 2}
\left( {2b\over n^2}\cos {nx\over 2} +
{2\over n} \sin {nx\over 2}\right)/F(x) 
\eqno(3.61)$$
is now the only periodic one among the expressions in (3.6).
This vector field has $2n$ {\it simple zeros} 
over the period $0\leq x < 2\pi$, 
and therefore we can identify $G[L_{n,b}]$  in terms
of the list in (3.9) as   
$$
G[L_{n,b}]\simeq T_{n,\Delta}.
\eqno(3.62{\rm a})$$
Using the explicit expression  in (3.58), 
the relation between the monodromy invariant $b$ and 
the invariant [\Ki,\Wi] $\Delta$ can be obtained from the 
definition [\Wi] of  $\Delta$ as 
$$
\Delta = 4\pi b.
\eqno(3.62{\rm b})$$
Observe from (3.59) that  $L_{n,b}$  
is periodic with period $2\pi\over n$.
Thus $G[L_{n,b}]\simeq T_{n,4\pi b}$ contains 
the  cyclic group ${\bf Z}_n$ of rigid rotations on $S^1$
by multiples of the angle $2\pi\over n$.
The structure of the  little group is further clarified in Appendix C.
We find that up to isomorphism 
$$
G[L_{n,b}]= {\bf R}_+ \times {\bf Z}_n 
\quad\hbox{and}\quad
\tilde G[L_{n,b}] = {\bf R}_+ \times {\bf T}_{{2\pi \over n}}
\eqno(3.63)$$
where ${\bf R}_+$ is the multiplicative group of positive real numbers
and ${\bf T}_{{2\pi \over n}}$ is the group of translations on ${\bf R}$
by multiples of ${2\pi \over n}$.
This is consistent with the fact that $\gamma$ in (3.18)
is a surjective homomorphism onto 
$G[B_\eta(b)]\simeq {\bf R}_+ \times {\bf Z}_2$.
The structure of $T_{n,\Delta}\simeq G[L_{n,b}]$ 
has not been fully clarified in Refs.~[\Ki,\Wi],
and our result (3.63) is in conflict with a claim of Ref.~[\G] 
on this point.

To summarize, we have established  the identification
$$
{\cal B}_n(b)={\cal O}_{L_{n,b}} = {\rm Diff}_0(S^1)/G[L_{n,b}]
= {\rm Diff}_0(S^1)/ T_{n,4\pi b}
\eqno(3.64)$$
and proved that up to conformal
transformations $\Psi$ in (3.58) represents the solutions of Hill's 
equation with monodromy $B_\eta(b)$, $\eta=(-1)^n$ 
 along this Virasoro coadjoint orbit.

\medskip
\noindent
{\bf 3.6.\ Virasoro coadjoint orbits with parabolic monodromy}
\medskip

The discussion is similar to that in  the previous subsection.
For any $L$ for which  the monodromy matrix of Hill's equation 
belongs to the conjugacy class of $P_\eta^q$ in (3.25a), 
consider a solution vector $\Psi=\pmatrix{\psi_1 &\psi_2}$ 
whose monodromy matrix equals  $P_\eta^q$. 
It follows from (3.25b) that $\psi_2$ is now unique up to sign. 
Hence the discrete invariant $n(L)$ defined by (3.43) can 
again be attached  to the Virasoro coadjoint orbit through $L$,
and $\eta=(-1)^n$ holds. 
Below, we show that the invariants $n\in {\bf Z}_+$ 
and $q=\pm 1$ together provide a {\it complete}  classification
of the Virasoro orbits in this case.
The orbit corresponding to a fixed value of $n\in {\bf Z}_+$
and $q\in \{ \pm 1\}$ will be denoted by ${\cal P}_n^q$. 
Sometimes we also write ${\cal P}^\pm_n$  to refer to  
${\cal P}_n^q$ for $q=\pm 1$, respectively.
The list of orbits turns out to be 
$$
{\cal P}_{0}^{+}, 
\qquad {\cal P}_n^q \quad  n\in {\bf N},\,\,\,\, q\in \{\pm 1\}.
\eqno(3.65)$$
If $n=0$, then  only the value $q=+1$ occurs.
We justify our claim by bringing any $L$ with
invariants $n$, $q$ to a standard form by a conformal transformation.

\smallskip
\noindent 
{\it 3.6.1.\ The case ${\cal P}_0^+$.}
Choosing any $\Psi$ with monodromy $P_1^q$,  
the smooth function $u={\psi_1\over \psi_2}$ now satisfies 
$$
u'={1\over \psi_2^2}>0
\quad\hbox{and}\quad
u(x+2\pi)=u(x) + q,
\eqno(3.66)$$
where $q=+1$ since $u$ is monotonically increasing.
This explains the remark after equation (3.65), and also implies  that
$\alpha := 2\pi u$ is an element of ${\widetilde{\rm Diff}}_0(S^1)$.
As a consequence of an
analogous  general property of $\psi_1\over \psi_2$,
one then obtains
$$
L= {\psi_2''\over \psi_2}=-{1\over 2}{u'''\over u'} +{3\over 4} 
{u''^2 \over u'^2} = S(u)=S(\alpha),
\eqno(3.67)$$
which means by (3.2) that $L$ is on the coadjoint orbit of
$$
L_0(x) := 0.
\eqno(3.68)$$
It is clear from (3.6) that the Lie algebra 
of the the little group $G[L_0]\subset {\rm Diff}_0(S^1)$ 
is spanned by
the vector field ${\partial \over \partial x}$, and $G[L_0]=S^1$ in fact.
At $L_0$,  two conformally nonequivalent 
solution vectors  of Hill's equation are 
given by $\Psi_0^\pm=\pmatrix{\psi_{1,0}^\pm &\psi_{2,0}^\pm}$: 
$$\eqalign{
\psi_{1,0}^\pm(x):&=\pm {x \over \sqrt{2\pi}},\cr
\psi_{2,0}^\pm(x):&=\pm \sqrt{2\pi}.\cr}
\eqno(3.69)$$
One readily confirms that the image of the homomorphism 
$\gamma: \tilde G[L_0]\rightarrow  G[P_1^1]$ defined by (3.18)
is the connected component of $G[P_1^1]$.
Thus $\Psi_{0}^\pm$ represent  the conformally nonequivalent 
solutions of Hill's equation with monodromy $P_1^1$ along the coadjoint
 orbit 
$$
{\cal P}_0^{+}={\cal O}_{L_0}=
{\rm Diff}_0(S^1)/S^1
={\widetilde{\rm Diff}}_0(S^1)/{\bf R}.
\eqno(3.70)$$

\smallskip
\noindent 
{\it 3.6.2.\ The case ${\cal P}_n^q$  for $n\in {\bf N}$.}
For any $q\in \{\pm 1\}$, $n\in {\bf N}$ and
 $\eta =(-1)^n$,
define $\Psi_{n,q}=\pmatrix{\psi_{1,n,q}&\psi_{2,n,q}}$ by 
$$\eqalign{
\psi_{1,n,q}(x):&={1\over \sqrt{H(x)}}
\left({qx\over 2\pi} \sin {nx\over 2} - 
{2\over n} \cos {nx\over 2}\right)\cr
\psi_{2,n,q}(x):&={1\over \sqrt{H(x)}}
\sin {nx\over 2}\cr}
\eqno(3.71{\rm a})$$
where  
$$
H(x)=H_{n,q}(x):=1 +{q\over 2\pi} \sin^2 {nx\over 2}.
\eqno(3.71{\rm b})$$
This is a solution vector of Hill's equation with
monodromy $P_\eta^q$ and $n$ zeros of $\psi_2$ at
$$
L_{n,q}={n^2 \over 2H} -
{3n^2(1+{q\over 2\pi})\over 4H^2}.
\eqno(3.72)$$
The proof in Appendix C shows that 
any Virasoro density $L$ of the same
monodromy type and discrete invariant $n(L)$  lies
on the coadjoint orbit of $L_{n,q}$, i.e., we have   
$$
{\cal P}_n^{q} ={\cal O}_{L_{n,q}}=
{\rm Diff}_0(S^1)/G[L_{n,q}].
\eqno(3.73)$$
The Lie algebra of $G[L_{n,q}]$ is spanned
by the vector field
$$
\xi(x){\partial \over \partial x},
\qquad
\xi(x)=\psi^2_{2,n,q}(x).
\eqno(3.74)$$
Since $\xi(x)$ has $n$ {\it double} zeros,  we can 
identify the little group in terms of the list (3.9) as 
$$
G[L_{n,q}]\simeq {\tilde T}_{n,{\rm sign}(-  q) },
\eqno(3.75)$$
where the flip of sign is merely our convention
for defining ${\tilde T}_{n,\pm }$.
Similarly to  $G[L_{n,b}]$ in (3.63),  $G[L_{n,q}]$ 
has the structure ${\bf R}_+\times {\bf Z}_n$ 
and  (3.18) gives a 
surjective homomorphism. 
The solution $\Psi_{n,q}$ in (3.71) is conformally equivalent 
to  any solution vector  of Hill's equation that has 
the same monodromy matrix $P_\eta^q$, $\eta=(-1)^n$  and 
discrete invariant $n$.

\medskip
\noindent
{\bf 3.7.\ Summary of the Virasoro coadjoint orbits}
\medskip

In this section, we described the coadjoint orbits of
the Virasoro algebra by exhibiting  representatives
for all  of them, and  presented an explicit list for 
the conformally nonequivalent solutions of Hill's equation.
We also provided the link between our 
description of the coadjoint orbits and the one  based
on the inspection of the possible little groups [\Ki,\Wi]
of the Virasoro densities.
In our notation, the set of orbits is given by 
$$
{\cal C}(\nu),
\quad
{\cal B}_0(b),
\quad
{\cal P}_0^+;
\quad\qquad
{\cal E}_n;
\quad\qquad
{\cal B}_n(b);
\quad\qquad
{\cal P}_n^\pm
\qquad (n\in {\bf N},\quad  b>0),
\eqno(3.76)$$
where we have arranged the orbits
into four families  according to the
type of the little group in (3.9). 
Ref.~[\GRS] contains related (but incomplete)
results on the connection between the monodromy invariant (3.21) 
and the type of the little group $G[L]\subset {\rm Diff}_0(S^1)$.
Observe that the first two families in (3.76) 
consist of the orbits ${\cal O}_L$ possessing a
constant representative $L=\Lambda$ for some $\Lambda\in {\bf R}$.

As a mnemonic,  we find it useful to associate the set 
of Virasoro orbits in (3.76) with the points 
of a `comb-like' figure, which we colloquially refer to as 
the `Vircomb' (see Figure 1).
To explain the shape of the Vircomb,  recall that 
the ${\cal B}_n(b)$ representatives in  (3.58)  become the 
${\cal E}_n$  ones in (3.42) as $b\rightarrow 0$.
As for the  ${\cal P}_n^q$ orbits, they are also  
deformations of the ${\cal E}_n$ orbits 
in a sense [\Wi].
To illustrate this, consider for example 
the  one parameter curve of  Virasoro densities
$L_{n,q;a}$ and corresponding solution vectors 
${\widehat  \Psi}_{n,q;a}$ given as follows: 
$$
L_{n,q;a}:={n^2 a^2 \over 2H_{n,q;a}}
-{3n^2 a^2 (a^2 + {q\over 2\pi})\over 4 H^2_{n,q;a}}
\quad\hbox{with}\quad 
H_{n,q;a}:=a^2 +{q\over 2\pi} \sin^2 {nx\over 2} 
\eqno(3.77{\rm a})$$ 
and  $\widehat{\Psi}_{n,q;a}= \Psi_{n,q;a} A(a)$,  where 
${\Psi}_{n,q;a}=\pmatrix{\psi_{1,n,q;a}&\psi_{2,n,q;a}}$ with 
$$\eqalign{
\psi_{1,n,q;a}(x):&={1\over \sqrt{H_{n,q;a}(x)}}
\left({qx\over 2\pi} \sin {nx\over 2} - 
{2a^2\over n} \cos {nx\over 2}\right)\cr
\psi_{2,n,q;a}(x):&={1\over 
\sqrt{H_{n,q;a}(x)}}\sin {nx\over 2}\cr}
\eqno(3.77{\rm b})$$
and 
$$
A(a):=\pmatrix{0&-{1\over a}\sqrt{n\over 2}\cr
a\sqrt{2\over n}&0\cr}.
\eqno(3.77{\rm c})$$
The parameter $a$ runs either as $a>0$ or as $a>1/\sqrt{2\pi}$
depending on whether $q=+1$ or $q=-1$, respectively. 
One can check that 
the  monodromy matrix of ${\widehat  \Psi}_{n,q;a}$
belongs to the class of $P_\eta^q$ 
for any finite $a$ and the discrete invariant takes the 
value $n$, which ensure that the curve $L_{n,q;a}$ 
lies on the orbit ${\cal P}_n^q$.
The point we wish to make  is that
$L_{n,q;a}$ has $L_n=-{n^2\over 4}$ 
as its $a\rightarrow \infty$ limit,
and the solution vector 
$\widehat{\Psi}_{n,q;a}$
becomes the ${\cal E}_n$ representative 
$\Psi_n$  (3.42) in this limit.
This curve will  be used in an argument later, too.

\bigskip

\input epsf
\hskip-1.5truecm
\epsfxsize=14.0cm
\epsfbox{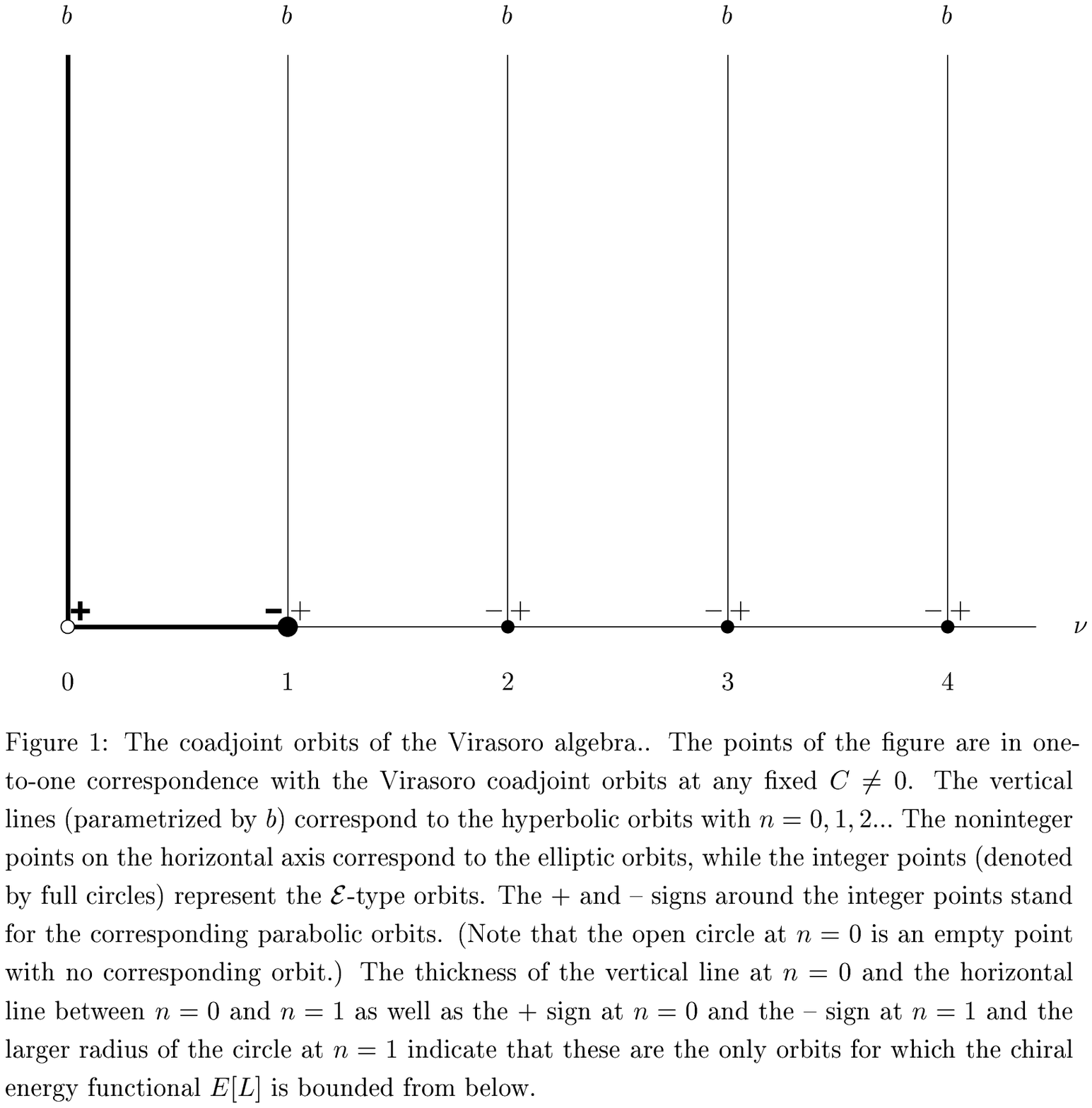}
\vskip4truecm

\bigskip\medskip

The approach whereby we derived the list in (3.76) 
is essentially an elementary 
version and elaboration of the
approach  followed in the  previous papers 
[\LP,\Seg,\OK], where the classification of the Virasoro orbits 
was also presented as
a refinement of the rough classification defined by the monodromy 
invariant (3.21).
We now give a more detailed comparison with the
results of Lazutkin and Pankratova [\LP]. 
These authors first define the following decomposition
of the set $X$ of the Hill's equations in (3.1):
$$
X=X^{\rm stable} \cup X^{\rm unstable} \cup X^{\rm semistable}\,,
\eqno(3.78)$$
where $X^{\rm stable}$ consists of the equations that possess only 
bounded solutions, $X^{\rm unstable}$ the equations whose nontrivial
solutions are all unbounded, and $X^{\rm semistable}$ is the rest.
In our notation, $X^{\rm stable}$ corresponds to 
elliptic and $E$ type monodromies, 
$X^{\rm unstable}$ corresponds to 
hyperbolic  monodromy,
and $X^{\rm semistable}$ corresponds to parabolic monodromy.
They next consider a second decomposition 
$X=X^{\rm nonosc} \cup X^{\rm osc}$, where $X^{\rm osc}$ 
contains the equations whose real solutions have infinitely many 
zeros both for positive and negative $x$.
In terms of (3.76),  we can identify these sets as 
$$
X^{\rm nonosc}=\{ {\cal P}_0^+,\ {\cal B}_0(b)\}
\quad\hbox{and}\quad
X^{\rm osc}=\{ {\cal C}(\nu),\ {\cal E}_n, \ {\cal B}_n(b),\ 
{\cal P}_n^\pm\,\}.
\eqno(3.79)$$
In effect, the elements of $X^{\rm osc}$ are labelled in Ref.~[\LP]
by an invariant $\theta(L)$ together with the monodromy class
in $PSL(2,\R)$.
The definition of the Lazutkin-Pankratova invariant $\theta(L)$, 
which unifies our invariants $\nu(L)$ and $n(L)$ that we defined
in a case by case manner, is particularly elegant.
For this one uses the
{\it translation number} $\Theta(\alpha)$ associated to any 
$\alpha \in \widetilde{\rm Diff}_0(S^1)$ by 
$\Theta(\alpha):= \lim_{m\to \infty} {1\over 2\pi m} \alpha^m (x)$,
where $\alpha^m = \alpha \circ \alpha^{m-1}$.
According to Poincar\'e, the limit exists and is independent of $x$.
Then $\theta(L):=\Theta(\alpha_L)$, where $\alpha_L$ 
is defined as follows. Let $\psi^x$ be a nontrivial real  
solution of Hill's equation that vanishes at $x$,
and set $\alpha_L(x)$ to be the zero of 
$\psi^x$ which is next to  $x$ in the positive direction
(such a zero exists since $\psi^x$ is oscillating).
It is not difficult to verify that 
$$
\theta(L) = {1\over \nu} 
\quad\hbox{if}\quad L\in {\cal C}(\nu) 
\quad\hbox{and}\quad
\theta(L)={1\over n} 
\quad\hbox{if}\quad
L\in {\cal E}_n,\, {\cal B}_n(b),\, {\cal P}_n^\pm 
\,\,\, \hbox{for}\,\,\, n\in {\bf N}.
\eqno(3.80)$$
This completes the correspondence between our classification 
and that in Ref.~[\LP].
We note that Lazutkin and Pankratova did not give 
explicit representatives for the solutions of (3.1) in the 
nontrivial cases ${\cal P}_n^\pm$, ${\cal B}_n(b)$, and
it is not clear [\Seg] whether 
their suggested representative Virasoro densities 
are correct in these cases or not.

\bigskip 
\centerline{\bf 4.~The energy functional on the coadjoint orbits}
\medskip

In the previous section we described the coadjoint orbits
of the Virasoro algebra at central charge $C\neq 0$.
We now recall that Witten [\Wi] also investigated the behaviour of the 
zero mode functional $\L^0$ on the orbits.
An interesting question, 
motivated for example  by analogy with 
the representation theory of the Virasoro algebra, 
is  to find the list of orbits
on which $\L^0$ is bounded from below.
As was already mentioned,
the zero mode functional is given by 
$$
\L^0 = {C\over 6}\left( E[L] + {1\over 4}\right)
\quad\hbox{where}\quad  
E[L]={1\over 2\pi}\int_0^{2\pi} dx\, L(x) 
\eqno(4.1)$$
is the so called chiral energy functional.
On the orbit ${\cal O}_L$ of $L$ fixed,
it is further convenient to define
$$
E_L[\alpha]:=E[L^\alpha],
\qquad \alpha\in \widetilde{\rm Diff}_0(S^1).
\eqno(4.2)$$
This functional is unbounded from above on every 
orbit (as is easily seen from (4.5)),
and hence $\L^0$ is bounded from below
if and only if $E_L[\alpha]$ is bounded from
below and $C>0$.
In fact, the orbits on which the chiral energy  functional 
is bounded from below are: 
$$
{\cal B}_0(b),\qquad
{\cal P}_0^+,\qquad
{\cal C}({\nu})
\quad\hbox{for}\quad 0<\nu<1,
\qquad
{\cal E}_1,
\qquad
{\cal P}_1^-,
\eqno(4.3)$$
see Figure 1.
Except for ${\cal P}_1^-$, these orbits contain a constant representative,
$$
L=\Lambda\quad\hbox{for}\quad \Lambda\geq  -{1\over 4},
\eqno(4.4)$$
and  the chiral energy has a global minimum
on the orbit taken precisely at the constant representative.
This representative may be called a `classical highest weight state' of the
Virasoro algebra if  $C>0$,
since then $\L^0$ is also bounded from below by 
its  global minimum  taken at the classical
highest weight state.
The  global minimum  of $\L^0$  is positive except for the
 `vacuum orbit' ${\cal E}_1$, where it takes the value zero.
On the orbit ${\cal P}_1^-$ with $C>0$,
the zero mode functional $\L^0$ has again the greatest lower  bound given 
by zero, but this  value is not taken on the orbit.
These results were all described in Ref.~[\Wi],
but in some cases (the case of ${\cal P}_1^-$ 
and the question of global minimum at $L=\Lambda$ in (4.4))  
no attempt was made to prove them.
Our aim below is to present a complete, elementary proof.
We shall proceed in  a case by case manner.

\medskip
\noindent
{\bf 4.1.~The case of the vacuum orbit ${\cal E}_1$}
\medskip
 
As a preparation, let us spell out (4.2) as 
$$
E_L[\alpha]={1\over 2\pi} \int_0^{2\pi} dy\,\left\{ \alpha'^2(y)L(\alpha(y))
+{1\over 4}{\alpha''^2(y)\over \alpha'^2(y)}\right\},
\eqno(4.5)$$
where we used
$$
S(\alpha)=
-{1\over 2}{\alpha'''\over \alpha'} +{3\over 4}
{\alpha''^2\over \alpha'^2}
=-{1\over 2}\left({\alpha''\over \alpha'}\right)' +
{1\over 4}{\alpha''^2\over \alpha'^2}.
\eqno(4.6)$$
Introducing the new integration variable
$$
y=\alpha^{-1}(x),
\quad
dy={dx\over q(x)}
\quad\hbox{with} \quad
q = {1\over (\alpha^{-1})'}=\alpha'\circ \alpha^{-1}
\eqno(4.7)$$
eq.~(4.5) becomes
$$
E_L[\alpha]={1\over 2\pi} \int_0^{2\pi} dx\,\left\{
q(x)L(x) +{1\over 4} {q'^2(x)\over q(x)}\right\}.
\eqno(4.8)$$
{}From its definition above,
$q$ is a smooth, $2\pi$-periodic function satisfying 
$$
q>0,
\qquad
{1\over 2\pi} \int_0^{2\pi} {dx\over q(x)}=1.
\eqno(4.9)$$
Conversely, any $2\pi$-periodic function $q$ subject to (4.9) 
determines a conformal transformation
$\alpha^{-1}$, which is unique up to translations.
Therefore we may regard $q$ with these properties
as the independent
variable in our extremum problem.
(When doing so, we will denote the chiral energy (4.8) as $E_L[q]$.)
Note also that 
if $\alpha\in \widetilde{\rm Diff}_0(S^1)$ is the lift
of a M\"obius transformation on $S^1$,
$$
e^{ix} \mapsto {a e^{ix}+\bar b\over b e^{ix}+\bar a},
\eqno(4.10)$$
where
$$
\pmatrix{a&\bar b\cr  b&\bar a\cr}\in SU(1,1),
\quad\hbox{i.e.}\quad
\vert a\vert^2-\vert b\vert^2=1,
\eqno(4.11)$$
then the corresponding function $q=1/(\alpha^{-1})'$ has the form
$$
q(x)=\sqrt{1+\mu^2} -\mu \cos(x+x_0)
\quad
\hbox{with}
\quad
\mu=2\vert a b \vert \geq 0,
\quad
x_0=\arg b - \arg a.
\eqno(4.12)$$

With this preparation in hand, we are ready to prove that on the orbit
$$
{\cal E}_1={\cal O}_{L_1}
\qquad
L_1=-{1\over 4},
\eqno(4.13)$$
the chiral energy functional 
$$
E_{L_1}[q]={1\over 8\pi} \int_0^{2\pi}dx\,\left\{
{q'^2\over q}-q\right\}
\eqno(4.14)$$
has the global mininum $-1/4$, and the minimum is taken at those
functions $q(x)$ that are of the form (4.12).
Since the M\"obius transformations leave $L_1$ invariant, the second 
statement means that the global minimum of the chiral energy 
on ${\cal E}_1$ is precisely at the `classical vacuum state' $L_1$. 
We start by integrating the inequality
$$
{1\over2\pi}\Bigg(
{|q^\prime|\over\sqrt{q}}-\sqrt{m+M-q-{mM\over q}}\,\Bigg)^2
\geq0
\eqno(4.15)$$
which, using the definition (4.14) and the constraint (4.9), gives
$$
4E_{L_1}[q]\geq-m-M+mM+{1\over\pi}\,\int_0^{2\pi}\,dx\,
{|q^\prime|\over q}\sqrt{\Big({M-m\over2}\Big)^2-
\Big(q-{M+m\over2}\Big)^2}\,.
\eqno(4.16)$$
Here $m$ and $M$ are the absolute minimum and maximum of the 
function $q(x)$, respectively. 

If there were no absolute value sign in the integrand in (4.16), we could
change the integration variable from $x$ to $q$. We can still do this
piecewise, between two consecutive local extrema of the function $q(x)$,
since within such an interval the sign of $q^\prime$ remains constant. 
We can now express the right hand side of (4.16) with the help of the
primitive function of the integrand,
$$
\cF(q)=\int_m^q\,{dz\over z}\,\sqrt{\Big({M-m\over2}\Big)^2-\Big(
       z-{M+m\over2}\Big)^2}\,.
\eqno(4.17)
$$
Let us look at this in detail for the case illustrated by Figure 2. 
(The general case can be treated similarly.) In this case we have
$$\eqalignno{
&4E_{L_1}[q]\geq -m-M+mM+{2\over\pi}
\Big[\cF(q_1)+\cF(q_3)+\cF(q_5)-\cF(q_2)-\cF(q_4)-\cF(q_6)\Big]\cr
&=-m-M+mM+{2\over\pi}\Big[\cF(M)-\cF(m)\Big]
+{2\over\pi}\Big[\cF(q_3)-\cF(q_4)\Big]
+{2\over\pi}\Big[\cF(q_1)-\cF(q_6)\Big]\cr
&\geq-m-M+mM+{2\over\pi}\Big[\cF(M)-\cF(m)\Big]\cr
&=mM-2\sqrt{mM}=
\big(\sqrt{mM}-1\big)^2-1\geq-1\,.&(4.18)\cr}
$$

\vskip-2truecm
\hskip1.0truecm
\epsfbox{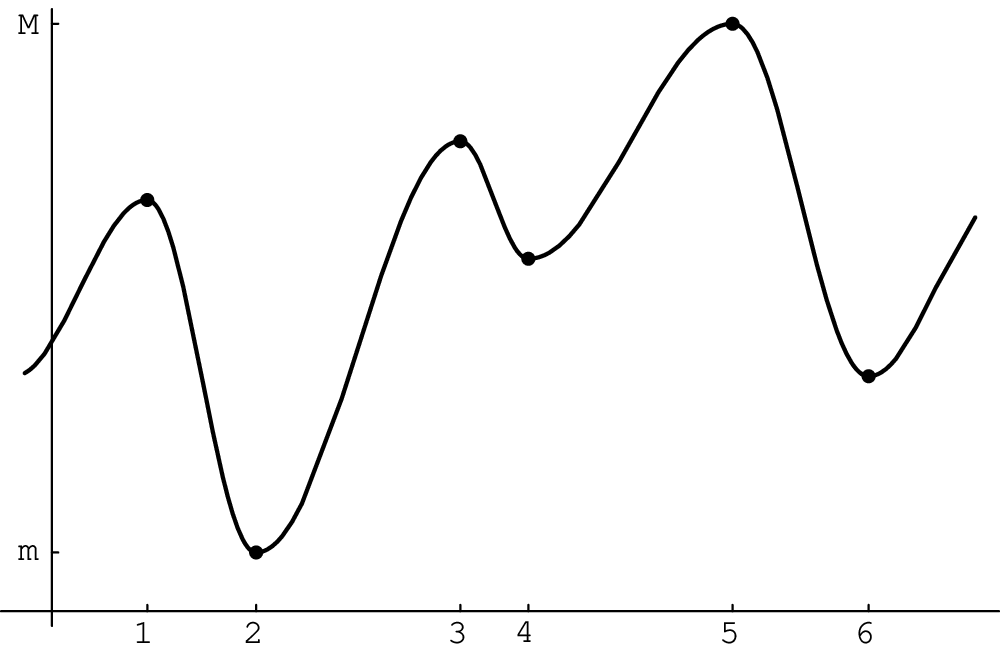}

\vskip-2truecm
\centerline{Figure 2:  \ \ An example of the function $q(x)$}
\bigskip
\noindent
This shows that $-1/4$ is a lower bound for the functional $E_{L_1}[q]$.
It is also clear from the derivation that this lower bound is saturated
only if the following three conditions are satisfied by the function $q(x)$:

\item{1)} It satisfies the differential equation
$$
{q^{\prime2}\over q}=m+M-q-{mM\over q}\,.
\eqno(4.19)
$$

\item{2)} In addition to the points where $q(x)$ takes its absolute
 minimum $m$ and maximum $M$ it has no other local extrema.

\item{3)} $$mM=1.\eqno(4.20)$$

The most general solution of the above conditions
is given by (4.12).
It is easy to check that (4.12) satisfies the 
requirements 1)-3) above. To see that it gives the
most general solution it is useful to rewrite (4.19)
in terms of $\beta:=\alpha^{-1}$,  $q=1/ \beta'$.
One obtains that the derivative of
(4.19) using (4.20) becomes the condition
$$
0=q'\left(L_1^\beta -L_1 \right).
\eqno(4.21)$$
It is well-known (and is shown in Appendix B) that the solutions of
$$
L_1^\beta = L_1
\eqno(4.22)$$
are the M\"obius transformations,
yielding (4.12) as remarked earlier.
Note that (4.22) may also be derived
directly from (4.5) as the
Euler-Lagrange equation of the functional
$E_{L_1}[\alpha]$.
This already implies that there is a unique  local
extremum  of the chiral energy at $L_1$,
but the above reasoning  also proves 
that this really is a global minimum, which is 
not obvious [\Wi].

\medskip
\noindent{\bf 4.2.~Arbitrary orbit with a constant representative}
\medskip
 
Consider the orbit ${\cal O}_\Lambda$ of a constant $L$,
$$
L=\Lambda,
\qquad \Lambda\in {\bf R}.
\eqno(4.23)$$
We wish to show that the behaviour of the chiral energy 
$E_\Lambda[\alpha]$ depends
on whether
$\Lambda \geq -{1\over 4}$
or
$\Lambda<-{1\over 4}$.
(Recall that ${\cal O}_\Lambda$ is of type ${\cal B}_0(b)$,
${\cal P}_0^+$, ${\cal C}(\nu)$ or ${\cal E}_n$ depending on the
particular value of the constant $\Lambda$.)
For any $\Lambda$, one has the identity:
$$
E_\Lambda[\alpha]=
{\Lambda+{1\over 4}\over 2\pi}\int_0^{2\pi}  dx\, \alpha'^2(x)+
E_{-{1\over 4}}[\alpha]
=\Lambda+ {\Lambda+{1\over 4}\over 2\pi}\int_0^{2\pi}  dx\,
\left(\alpha' -1\right)^2+
\left(E_{-{1\over 4}}[\alpha]+{1\over 4}\right).
\eqno(4.24)$$
If one now takes
$$
\Lambda>-{1\over 4},
\eqno(4.25)$$
then (4.24) and the result of the previous subsection
imply that
$$
E_\Lambda[\alpha]\geq \Lambda,
\eqno(4.26)$$
and equality holds if and only if $\alpha'=1$,
which holds precisely  for the elements of 
the little group of $L=\Lambda$.
This proves that in the case (4.25) the
chiral energy has a unique,
global minimum on the orbit ${\cal O}_\Lambda$ taken at $L=\Lambda$,
like for $\Lambda=-{1\over 4}$.
 
In contrast, if
$$
\Lambda<-{1\over 4},
\eqno(4.27)$$
then the chiral energy is not bounded from below on the orbit 
${\cal O}_\Lambda$.
To see this, consider the one parameter subgroup of the M\"obius group
given by
$$
\pmatrix{ \cosh {t}&  i\sinh {t}\cr
          -i\sinh{t}& \cosh {t}\cr}
\in SU(1,1),
\qquad
\forall\, t\in {\bf R},
\eqno(4.28)$$
for which  one has
$$
q_t(x)=\cosh 2t -  \sinh 2t   \sin x.
\eqno(4.29)$$
One then finds that 
$$
E_\Lambda[q_t]=-{1\over 4}+ \left(\Lambda
+{1\over 4}\right)\cosh 2t.
\eqno(4.30)$$
Since $t$ is arbitrary, the chiral energy is indeed not bounded
from below in the case (4.27).

\medskip
\noindent{\bf 4.3.~The orbits ${\cal P}_n^-$ for
$n\geq 2$ and  ${\cal P}_n^+$, ${\cal B}_n(b)$ for any $n\in {\bf N}$}
\medskip
 
We  wish to show
 that the chiral energy is not bounded from
below on these orbits.
First, consider the orbit ${\cal P}_n^-$, whose
representative $L_{n,-}$ is defined by (3.71-72) with $q=-1$.
One can verify the inequality
$$
L_{n,-}(x) < - {n^2\over 8}.
\eqno(4.31)$$
Therefore
$$
L_{n,-}(x) < -{1\over 2}
\quad\hbox{if}\quad
n\geq 2,
\eqno(4.32)$$
which implies by comparison with a constant $L=\Lambda$ in (4.27)
that the chiral energy is not bounded from
below in this case.
 
Second,  using the curve $L_{n,+;a}$  
of Virasoro densities on  the orbit ${\cal P}_n^+$
given by (3.77a) with $q=+1$,
one obtains  that 
$$
E[L_{n,+;a}]=-{n^2\over 8\sqrt{a^2 + 1/2\pi}}\left(
{3/2\pi +2a^2 \over a}\right).
\eqno(4.33)$$
This approaches  $-\infty$ as $a \rightarrow 0$,
settling the ${\cal P}_n^+$  case.
 
The ${\cal B}_n(b)$, $n\in {\bf N}$, case is dealt with 
by a similar computation using a one parameter curve 
on this orbit, given by
$$\eqalign{
L_{n,b;a}=&b^2 +{n^2 +4b^2\over 2F_{n,b;a}} -{3\over 4}
{n^2a^2\over F_{n,b;a}^2},\cr
F_{n,b;a}(x) =&a^2\cos^2{nx\over 2} +
\left( \sin {nx\over 2} +{2b\over n}\cos {nx\over 2}\right)^2,\cr}
\eqno(4.34)$$
where the parameter $a>0$ is arbitrary.
The corresponding solution $\Psi_a=\pmatrix{\psi_{1,a}&\psi_{2,a}}$
of Hill's equation is written as 
$$\eqalign{
\psi_{1,a}(x):&={e^{bx}\over \sqrt{F_a(x)}}\sqrt{n\over 2}
\left( \left[{2b\over n^2}+{a^2-1\over 2b}\right]
\cos {nx\over 2} +{2\over n} \sin {nx\over 2}\right)\cr
\psi_{2,a}(x):&={e^{-bx}\over \sqrt{F_a(x)}}\sqrt{2\over n}
\cos {nx\over 2}.\cr}
\eqno(4.35)$$
The existence  of  conformal
transformations `creating' this curve from the
`standard point' in  (3.58-59) (which corresponds to  $a=1$)
follows from the fact that (4.35) yields the same monodromy and 
discrete invariants as (3.58a).

Now it is easy to verify that
$$
E[L_{n,b;a}]=b^2 +{n^2 +4b^2\over 8a} - {3n^2 a\over 8},
\eqno(4.36)$$
which is not bounded from below as $a\rightarrow \infty$,
completing the proof of the above claim.

\medskip
\noindent 
{\bf 4.4.~The exceptional  orbit ${\cal P}_1^-$}
\medskip

In this subsection we prove that on the orbit 
${\cal P}_1^-$ the chiral energy functional $E[L]$  
has the largest lower bound $-1/4$, but this value is not reached.
The fact that makes ${\cal P}_1^-$ 
 exceptional is that for this orbit
(and only for this one among the ${\cal P}_n^q$, $n\neq 0$)
it is possible to find a number $\xi$ such that
$$
u(\xi)={1\over2}
\eqno(4.37)
$$
and
$$
\psi_1(x)>0 \qquad\qquad {\rm for}
\qquad\qquad \xi<x<\xi+2\pi\,,
\eqno(4.38)
$$
where $u(x)=\psi_1(x)/\psi_2(x)$, as in Appendix C. Indeed, we can 
always choose $\xi$ such that (4.37) is satisfied (since the function
$u(x)$ takes all values between two consecutive zeros of $\psi_2$) and
for which $\psi_2(\xi)$ is positive (since $\psi_2$ is antiperiodic).
But then (4.38) is also satisfied since $\psi_1(\xi)={1\over2}
\psi_2(\xi)>0$, $\psi_1(\xi+2\pi)=u(\xi+2\pi)\psi_2(
\xi+2\pi)=-{1\over2}\psi_2(\xi+2\pi)={1\over2}\psi_2(\xi)
=\psi_1(\xi)>0$ 
and $\psi_1$ cannot change sign between $\xi$ and $\xi+2\pi$.
The last statement holds 
since a possible zero of $\psi_1$ would be a zero of $u(x)$ as well,
but we know (see Appendix C) that $u(x)$ increases from 
${1\over2}$ to $+\infty$ as $x$ changes from $\xi$ to the 
singular point of $u(x)$ (the only one in this interval), and then
from $-\infty$ to $-{1\over2}$ as $x$ completes its period from
the singular point to $\xi+2\pi$.
Our proof is based on the above observation. 

We will  use the relations
$$
\psi_1^\prime(\xi+2\pi)=\psi_1^\prime(\xi)-
      {1\over\psi_1(\xi)}
\eqno(4.39)
$$ and
$$
\int_{\xi}^{\xi+2\pi}\,{dx\over\psi_1^2(x)}=
{-1\over u(\xi+2\pi)}+{1\over u(\xi)}=4\,,
\eqno(4.40)
$$
which also follow from (3.13-14) using the monodromy matrix 
$P^q_\eta$ in (3.25a) with $\eta=q=-1$.

As a consequence of the above, the smooth function $y(x):=
\psi_1(x+\xi+\pi)$ satisfies
$$
y(x) >0\quad\hbox{for}\quad
x\in [-\pi, \pi],
\qquad 
y(\pi)=y(-\pi)\,,\qquad y^\prime(\pi)=y^\prime(-\pi)-{1\over
y(\pi)}
\eqno(4.41)
$$
and
$$
\int_{-\pi}^\pi\,{dx\over y^2(x)}=4\,.
\eqno(4.42)
$$
We denote the absolute maximum of $y(x)$ in the interval $[-\pi,\pi]$
by $M$.

Now we express the chiral energy functional (4.1) in terms of $y(x)$ as
$$
E[y]={1\over2\pi}\int_{-\pi}^\pi\, dx\, L(x)={1\over2\pi}
\int_{-\pi}^\pi\, dx\, {y^{\prime\prime}\over y}={1\over2\pi}
\int_{-\pi}^\pi\, dx\, {y^{\prime2}\over y^2}-{1\over2\pi y^2(\pi)}\,.
\eqno(4.43)
$$
 
The following consideration is very similar to that of Sect.~4.1.,
so here we can be brief. We start by integrating the inequality
$$
{1\over2\pi}\Bigg({|y^\prime|\over y}-w\sqrt{{M^2\over y^2}-1}\Bigg)^2
\geq 0,
\eqno(4.44)
$$
where $w$ is a positive constant to be fixed later and obtain
$$
E[y]\geq-{1\over2\pi y^2(\pi)}-w^2\Bigg({2M^2\over\pi}-1\Bigg)+
{w\over\pi}\int_{-\pi}^\pi\, dx\,{|y^\prime|\over y^2}\sqrt{M^2-y^2}\,.
\eqno(4.45)
$$
Then, by an argument similar to that in Sect.~4.1., we can show that
$$
\int_{-\pi}^\pi\, dx\,{|y^\prime|\over y^2}\sqrt{M^2-y^2}\geq
2\,\int_{y(\pi)}^M \,{dy\over y^2}\sqrt{M^2-y^2}=2({\rm tg}\, p-p)\,,
\eqno(4.46)
$$
where equality holds only if $y(x)$ has a single local extremum in the
interval $[-\pi,\pi]$ (where it takes its absolute maximum $M$) and 
$p$ is introduced via the parametrization
$$
M={y(\pi)\over{\rm cos}\, p}\qquad\qquad\qquad 0<p<{\pi\over2}\,.
\eqno(4.47)
$$
So far we have shown that for any positive $w$
$$
E[y]\geq{-1\over2\pi y^2(\pi)} -Bw^2+2Aw\,,
\eqno(4.48)
$$
where the positive constants $A$, $B$ are
$$
A={1\over\pi}\Big({\rm tg}\, p-p\Big)\qquad\qquad\qquad B={2M^2\over\pi}-1
\eqno(4.49)
$$
(positivity of $B$ is a consequence of the constraint (4.42)).
Choosing  $w:=A/B$  yields 
$$
E[y]\geq {A^2\over B}-{1\over2\pi y^2(\pi)}\,.
\eqno(4.50)
$$
We now parametrize $y(\pi)$ as
$$
y^2(\pi)={\pi\over2}{{\rm sin}\,2r\over2r}\qquad\qquad\qquad
 0<r<{\pi\over2}\,.
\eqno(4.51)
$$
This parametrization is not possible if $y^2(\pi)\geq\pi/2$, but
in this case $E[y]\geq-1/\pi^2$ manifestly. In this parametrization
(4.50) can be written as 
$$
E[y]\geq f(p,r)=-{1\over\pi^2}{r\over{\rm sin}\, r\,{\rm cos}\, r}+
{r\over\pi^2}{({\rm sin}\, p-p\,{\rm cos}\, p)^2\over
{\rm sin}\, r\,{\rm cos}\, r-r{\rm cos}^2p}\,.
\eqno(4.52)
$$
It is  elementary to show that for $0<p,r<\pi/2$
$$
f(p,r)\geq f(r,r)=-{1\over\pi^2}\Big(r^2+r\,{\rm ctg}\, r\Big)>
-{1\over4}\,,
\eqno(4.53)$$
establishing the strict lower bound $-1/4$.

Finally, it is possible to compute the value of the 
chiral energy functional 
for the curve (3.77a) on ${\cal P}_1^-$ 
 explicitly. We find that  $E\rightarrow-1/4$ as $a\rightarrow\infty$
along this curve. 
Thus we have proven that $-1/4$ is the largest lower bound but
it is not taken on the orbit ${\cal P}_1^-$.
This is quite an intriguing orbit since (by (4.1), if $C>0$)  
the corresponding lower bound of the Virasoro zero mode $\L^0$ is zero.

\vfill\eject

\bigskip
\centerline{\bf 5.~The solutions of the global Liouville equation}
\medskip

The smooth,  periodic solutions of the global Liouville equation (1.1)
that we are interested in have two important characteristics.
The first is the topological type defined by the number of 
zeros of $Q(\tau, \sigma)$ for $0\leq \sigma < 4\pi$ at any fixed $\tau$.
This number, $N$, is even on account of the $4\pi$-periodicity 
and the fact that $Q$ cannot have a double zero in $\sigma$. 
The second is the `Virasoro orbit type' ${\cal O}_+\times {\cal O}_-$
attached to $Q$ by the orbits ${\cal O}_\pm$ of the Virasoro densities 
$L_\pm =\pa_\pm^2 Q /Q$.
This characteristic is a refinement of the monodromy type
of the solutions $\Psi_\pm$ of (1.3) that compose $Q$ according to (1.5).
Correspondingly,  we call  a solution of (1.1)  elliptic, 
hyperbolic etc if the solutions of the Hill's equations
(1.3) have elliptic, hyperbolic etc monodromy.
The main purpose of this section is to establish the precise 
relationship between the two characteristics.

We shall make use of the normal forms of Hill's equation 
to write down explicit representatives for all smooth, periodic  
solutions of (1.1) up to conformal transformations.
As explained in the Introduction, these representatives can be obtained
by taking $\Psi_+$ in (1.5) to run through the representative solutions
given in Section 3, and choosing $\Psi_-$  from an analogous
set of representatives in such a way to obey the monodromy constraint (1.7),
which ensures the periodicity of $Q$.
Inspection of the representative solutions  will then reveal 
the relation between the topological type and the Virasoro orbit type.
The results in  Section 4 will further permit to describe the behaviour of 
the energy functional
in the  various sectors of the global Liouville model.
According to eqs.~(2.22-23), the (total) energy of a solution $Q$ 
is given by 
$$
{\cal H}_{\rm WZ}[Q]=2\pi \kappa \left( E[L_+] + E[L_-]\right),
\eqno(5.1)$$ 
where $E[L_\pm]$ is the chiral energy at the respective Virasoro 
density $L_\pm$.
The energy is bounded from below on the conformal orbit
$\{  Q^{\alpha_+, \alpha_-}\}$ of a solution $Q$ (see (1.9))
if and only if the corresponding chiral energy functionals 
$E_{L_\pm}[\alpha_\pm]$ are both bounded from below and $\kappa >0$.
Hence, in the rest of this section we assume that $\kappa >0$.

Below, the number of zeros of $Q$ turns out to be a function of the
discrete parameters of the Virasoro orbits.
For elliptic orbits ${\cal C}(\nu_+)\times {\cal C}(\nu_-)$,
we find that $N=(\nu_+ +\nu_-)$, where 
the sum only depends on the discrete parameters
$d_\pm$ attached to $\nu_\pm$ by (3.33).
For ${\cal E}_n$ type orbits the same formula is valid in the limiting 
case of integral $\nu_\pm$.
Actually, the formula $N=(\nu_+ +\nu_-)$ can be used 
to describe most  other cases, too,  if we deform the representatives of the
Virasoro orbits according to Figure 1
and note  that as a topological 
invariant $N$ cannot depend on such deformations.
(It is for the same reason that $N$ cannot depend on the continuous
parameters of the monodromy.)
Consistently with the earlier results [\Fad,\Ald],
we shall see that the $N=0$ sector, which contains the regular 
solutions of the Liouville equation (1.2), 
is associated with the Virasoro orbits ${\cal B}_0(b)\times {\cal B}_0(b)$,
and it turns out that the  energy is  bounded from below in 
this sector only. 

In Ref.~[\Mar] it was argued that the N=2 sector is special in that it
contains a stable lowest energy state (the one with constant 
Virasoro densities) and is therefore a suitable starting point 
for quantization. As we show in Appendix D, the N=2 sector is connected
and the would-be vacuum state of Ref.~[\Mar] (which in our notation is 
given by eq.~(5.8) below) is only locally stable since it
is continuously connected to states whose energy is not bounded from below.

\medskip
\noindent 
{\bf 5.1.~Elliptic solutions}
\medskip

If $Q$ is a solution  so that $\Psi_+$ in (1.5)  has 
elliptic monodromy, then using the freedom (1.8)
we may assume that $M_{\Psi_+}$ is of the
standard form (3.23a) with some $\omega:=\omega_+$,
$$
M_{\Psi_+}=C(\omega_+),
\quad 0<\omega_+ <2\pi,
\quad
\omega_+ \neq\pi.
\eqno(5.2{\rm a})$$
Then (1.7) forces
$$
M_{\Psi_-}=C(\omega_-)
\quad
\hbox{with}
\quad
\omega_-=2\pi -\omega_+.
\eqno(5.2{\rm b})$$
Therefore the parameters $\nu_{\pm}$ of
the respective chiral and anti-chiral 
Virasoro orbits ${\cal C}(\nu_{\pm})$  take the form
$$
\nu_+={\omega_+\over \pi}+2d_+,
\qquad
\nu_-=\left(2-{\omega_+\over \pi}\right) +2d_-,
\qquad
d_\pm\in {\bf Z_+}\,,
\eqno(5.3{\rm a})$$
which fixes the  the allowed pairing 
$$
{\cal C}(\nu_+)\times {\cal C}(\nu_-).
\eqno(5.3{\rm b})$$
{}From the results of Section 3.3, a representative
solution $Q$ with this Virasoro orbit type can be given explicitly as
$$
Q(x^+,x^-) = {2\over \sqrt{\nu_+\nu_-}}\cos \theta,
\eqno(5.4{\rm a})$$
where
$$
\theta ={1\over 2}(\nu_+ x^+-\nu_- x^-)=
{1\over 2}\left({\omega_+\over\pi} +d_+-d_--1\right) \tau
+{1\over 2}\left(d_+ +d_- +1\right)\sigma.
\eqno(5.4{\rm b})$$
The Virasoro densities belonging to this solution
are
$L_\pm=-{1\over 4}\nu_\pm^2$,
and $Q^{\alpha_+,\alpha_-}$ in (1.9) is the general
solution associated with elliptic monodromy.
The number of zeros of such a solution over a $4\pi$ 
period  in $\sigma$ is
$$
N=2(d_+ + d_- +1)=\nu_+ + \nu_-.
\eqno(5.5)$$
In particular, $N\neq 0$.

On the orbit
of 
the conformal group 
$\widetilde {\rm Diff}_0(S^1)\times \widetilde {\rm Diff}_0(S^1)$ 
through $Q$ in (5.4) the energy is not bounded  from below.
This follows from the results in the preceding  section since
either $\nu_+$ or $\nu_-$ is greater than $1$ by (5.3a) .

\medskip
\noindent
{\bf 5.2.~Solutions associated with monodromy type $E_\eta$} 
\medskip

By (1.7), the list of possible 
Virasoro orbit types is in this case
$$
{\cal E}_{n_+} \times {\cal E}_{n_-}
\quad
\hbox{with}\quad
 n_\pm \in {\bf N},\quad
(n_++n_-)\in 2{\bf N}.
\eqno(5.6)$$
For any such orbit,  there is a unique solution of (1.1) up to conformal 
transformations: 
$$
Q(x^+,x^-)={2\over \sqrt{n_+n_-}}\cos{(n_+x^+ -n_-x^-)\over 2},
\eqno(5.7)$$
whose topological type is given by $N=(n_++n_-)$.
Of particular interest is the $N=2$ representative solution,
$$
Q(x^+,x^-)= 2 \cos\left({1\over 2}\sigma\right).
\eqno(5.8)$$
This is the only one among the $E_\eta$-type 
solutions such that on the orbit
$\{Q^{\alpha_+,\alpha_-}\}$ 
of $\widetilde {\rm Diff}_0(S^1)\times 
\widetilde {\rm Diff}_0(S^1)$ 
the  energy is {\it bounded} from below.
The static solution  $Q$ in (5.8) may be thought of as the 
classical analogue of a vacuum state in a conformal field theory,
since the Virasoro zero modes $\L^0_\pm$ (4.1) are equal to zero at 
$Q$ and are bounded from below on the associated coadjoint orbits.
However, unlike in standard conformal field theory,
the energy is not bounded from below in the global Liouville model (1.1).

\medskip
\noindent
{\bf 5.3.~Hyperbolic solutions}
\medskip

When constructing the hyperbolic solutions of (1.1) by (1.5), 
we may assume that the  chiral monodromy matrix $M_{\Psi_+}$ 
is of the standard form $B_\eta(b)$ in (3.22a), and then 
(1.7) requires the anti-chiral monodromy
matrix $M_{\Psi_-}$ to coincide with the chiral one.
This means that the set of hyperbolic solutions 
is a union over the Virasoro orbit types
$$
{\cal B}_{n_+}(b)\times {\cal B}_{n_-}(b)
\quad
\hbox{where}
\quad
b>0,\quad (-1)^{n_++n_-}=1.
\eqno(5.9)$$
We shall verify that the 
topological type of these solutions is given by 
$$
N=(n_++n_-).
\eqno(5.10)$$
On a set of solutions whose Virasoro orbit type involves
${\cal B}_n(b)$ for some $n>0$,
the energy is not bounded from below.
Next we elaborate the simplest case $n_\pm=0$, for which the
energy is bounded from below.

\smallskip
\noindent
{\it 5.3.1.~Solutions with Virasoro orbit type 
${\cal B}_0(b)\times {\cal B}_0(b)$.}
For fixed $b>0$, the conformally nonequivalent solutions
of Hill's equation are given by (3.55).
Applying this to the chiral and anti-chiral
equations in (1.3), we obtain from (1.5) the following
two  conformally nonequivalent solutions of the
global Liouville equation:
$$
Q_\pm (x^+,x^-)=\pm {1\over b}\cosh b(x^++x^-)=\pm {1\over b}\cosh b\tau. 
\eqno(5.11)$$
As they are $\sigma$ independent, 
these solutions have no zeros, $N=0$.
We shall see that all other Virasoro orbit types belong to $N\neq 0$.

The global Liouville equation (1.1) is invariant
under $Q\mapsto -Q$,  and the topological sector $N=0$ 
further decomposes into two disconnected 
subsectors according to the sign of $Q$.
Clearly,  both these subsectors  correspond  to the
regular solutions  of the Liouville equation(1.2) by the substitutions 
$Q=\pm e^{-{1\over 2}\varphi }$.

It is interesting to recover 
the standard form of the solutions  of (1.2) from 
$Q_+^{\alpha_+,\alpha_-}$  with $Q_+$ in (5.11),
which yields 
$$
e^{-{1\over 2}\varphi(x^+,x^-)}=
{1\over b \sqrt{\pa_+\alpha_+(x^+) \pa_-\alpha_-(x^-)}} 
\cosh b\left(\alpha_+(x^+)+\alpha_-(x^-)\right),
\eqno(5.12)$$
where  $\alpha_+$, $\alpha_-$ now satisfy (1.10),
ensuring the periodicity and the smoothness of $\varphi$.
This can be rewritten in Liouville's form [\Liu] 
$$
\varphi(x^+,x^-)=\ln \left( 
{\pa_+ A(x^+) \pa_- A_-(x^-) \over 
\left(A_-(x^-)-A_+(x^+)\right)^2}\right)
\eqno(5.13)$$
by putting
$$
A_+(x^+):=-e^{-2b\alpha_+(x^+)},
\qquad
A_-(x^-):= e^{2b\alpha_-(x^-)}.
\eqno(5.14)$$
Of course,  if we `forget' these definitions  of $A_\pm$,
then we could also describe the singular solutions of (1.2) by (5.13).
However, it appears more economical and transparent to do so in 
terms of the variable $Q$ by means of our group theoretic analysis.

\smallskip\noindent
{\it 5.3.2.~Solutions of type ${\cal B}_0(b) \times {\cal B}_{2n}(b)$ 
or ${\cal B}_{2n}(b) \times {\cal B}_{0}(b)$ for $n\in {\bf N}$.}
By the analysis in Section 3.5, 
one needs  a singe  representative solution $Q$  of type 
${\cal B}_0(b)\times {\cal B}_{2n}(b)$, e.g.,
$$
Q(x^+,x^-)={1\over\sqrt{2nbF_{2n,b}(x^-)}}\left\{ e^{-b\tau}\cos nx^-
+e^{b\tau}\left( \sin nx^- +{b\over 2n}\cos nx^-\right)\right\},
\eqno(5.15)$$
where $F_{2n,b}$ is defined in (3.58).
It follows  that
$Q\vert_{\tau=0}=0$ if and only if  
$\,{\rm tg}\, {n\sigma\over 2}$=${ n+b\over  n }$,
which implies that $N=2n$.
In the case 
${\cal B}_{2n}(b)\times {\cal B}_0(b)$,
a representative solution is obtained from (5.15) using 
the symmetry of the global Liouville equation under the interchange of
the light cone coordinates, whereby 
$Q\mapsto \tilde Q$ with ${\tilde Q}(x^+,x^-):=Q(x^-,x^+)$.

\smallskip\noindent
{\it 5.3.3.~Virasoro orbit type
${\cal B}_{n_+}(b)\times {\cal B}_{n_-}(b)$ for $n_+n_-\neq 0$.}
Using the solution (3.58) of Hill's equation, 
we can now write down the representative
solution of the global Liouville equation as 
$$
Q(x^+,x^-) = 2 { e^{-b\tau}  \cos {n_+x^+\over 2}\cos {n_-x^-\over 2}
+ e^{-b\tau} X
\over 
\sqrt{n_+n_-F_{n_+,b}(x^+)F_{n_-,b}(x^-)}}
\eqno(5.16)$$
$$
X=\left(\sin {n_+x^+\over 2} +{b\over n_+}
\cos {n_+x^+\over 2}\right)
\left(\sin {n_-x^-\over 2} +{b\over n_-}\cos {n_-x^-\over 2}\right).
$$
Either directly comparing (5.7) and (5.16),
or by recalling a remark after (3.59),
we see that  this hyperbolic solution is a smooth
deformation of the ${\cal E}_{n_+}\times {\cal E}_{n_-}$ type solution 
(5.7), to which it reduces in the $b=0$ limit. 
Therefore $Q$ in (5.16) has the same number of zeros  as $Q$ in (5.7), 
namely, $N=(n_++n_-)\in 2{\bf N}$.

\medskip\noindent
{\bf 5.4.~Parabolic solutions}
\medskip

The representatives $P_\eta^q$
of the parabolic  conjugacy classes in (3.25a) obey 
$$
\left(P_\eta^q\right)^T =w^{-1} P^{-q}_\eta w
\quad\hbox{where}\quad 
w:=\pmatrix{0&-1\cr 1&0\cr}.
\eqno(5.17)$$
Hence in the parabolic case 
the  periodic solutions of (1.1) have Virasoro orbit types  
$$
{\cal P}_{n_+}^q \times {\cal P}^{-q}_{n_-}, 
\qquad
n_\pm \in {\bf Z_+},\quad
(n_++n_-)\in 2{\bf N}.
\eqno(5.18)$$
If we choose  $M_{\Psi_+}=P_\eta^q$ ($\eta=(-1)^{n_+}$) in the
construction (1.5), then $M_{\Psi_-}$ 
has to be chosen as $M_{\Psi_-}=w^{-1} P_\eta^{-q} w$. 
The solutions of Hill's equation with monodromy $M_{\Psi_-}$ 
can be obtained from those  in Section 3.6  by means of  
a transformation like in (3.15). 
Observe that the energy is not bounded from below 
in correspondence with any of the Virasoro orbit types in (5.18).
It turns out that the topological type of the solutions 
is given by  $N=(n_++n_-)$ similar to the hyperbolic case (5.10).

\smallskip\noindent
{\it 5.4.1.~Solutions of type 
${\cal P}_0^+ \times {\cal P}_{2n}^-$ or
${\cal P}_{2n}^- \times {\cal P}_{0}^+$ for $n\in {\bf N}$.}
It is clear from the study of Hill's equation  
in Section 3.6  that we may choose a single 
representative solution
of type  ${\cal P}_0^+ \times {\cal P}_{2n}^-$,  for example,
$$
Q(x^+,x^-)=\sqrt{2\pi\over H_{2n,-}(x^-)}
\left({\tau\over 2\pi} \sin nx^- +{1\over n}\cos nx^-\right),
\eqno(5.19)$$
with $H_{2n,-}$ defined in (3.71b).
Setting $\tau=0$,  we see that the number of zeros of $Q$ is  $2n$.
We can use  the symmetry of (1.1) under $x^\pm \mapsto x^{\mp}$  
to obtain a solution
of type ${\cal P}_{2n}^- \times {\cal P}_{0}^+$.

\smallskip\noindent
{\it 5.4.2.~Virasoro orbit type
${\cal P}_{n_+}^q\times {\cal P}_{n_-}^{-q}$ for $n_+n_-\neq 0$.}
Up to conformal transformations, the corresponding 
solution is given as follows:
$$
Q(x^+,x^-)={\left({2\over {n_-}}\sin {n_+x^+\over 2}\cos {n_-x^-\over 2}
 +{q \tau\over 2\pi}\sin {n_+x^+\over 2}\sin {n_-x^-\over 2}
-{2\over n_+}\cos {n_+x^+\over 2}\sin {n_-x^-\over 2}\right)
\over \sqrt{H_{n_+,q}(x^+)H_{n_-,-q}(x^-)}}.
\eqno(5.20)$$
At $\tau =0$,
this simplifies to the product of a factor without zeros and 
$$
Y(\sigma):=2n_-\sin {n_-\sigma\over 4}\cos {n_+\sigma\over 4}+
2n_+\sin {n_+\sigma\over 4}\cos {n_+\sigma\over 4}.
\eqno(5.21)$$
One can  check that $Y(\sigma) $ has $(n_++n_-)$ zeros 
for $0\leq \sigma <4\pi$. 

Alternatively, one may prove the relation  $N=(n_++n_-)$ 
using the curve  of parabolic solutions of Hill's equation given in (3.77),
which become the ${\cal E}_n$ 
representatives in the $a\rightarrow \infty$ limit.
Noting  that the topological type of the
corresponding $a$-dependent family of solutions of (1.1) is constant, 
one obtains $N=(n_++n_-)$ for the parabolic solutions 
of (1.1) as a consequence of the analogous relation in the 
$a\rightarrow \infty$ limit.

\smallskip

To summarize,  we enumerated the conformally nonequivalent,
smooth, periodic solutions of the global Liouville equation, 
and established the dependence of the topological type  $N$ on the
discrete parameters of the Virasoro coadjoint orbits.
Combining these results with the analysis of Section 4,
we may conclude that the energy 
${\cal H}_{\rm WZ}[Q]$ in (5.1) (with $\kappa >0$)
is bounded from below only in the
$N=0$ topological sector of the model.
For instance, if $N=2$ then we have solutions with Virasoro orbit type
${\cal E}_1 \times {\cal E}_1$ on which the energy is bounded from
below, and also of type ${\cal B}_1(b) \times {\cal B}_1(b)$,
as well as others, on which the energy is unbounded.
If $N=0$ then the sign of $Q$ is a further topological
invariant, but otherwise any two solutions with the same
number of zeros can be deformed into each other.
This can be seen with the aid of our explicit 
results on the solutions,
or can be established  by directly 
 inspecting the phase space of
the global Liouville model, the details are collected in Appendix D.

For completeness, let us finally comment on the present understanding of 
the role of the various classical solutions in the quantized Liouville 
field theory. First, the quantizations of Thorn et al [\THE] and Otto and 
Weigt [\OTW] deal with the regular sector, which contains a subset of the 
hyperbolic solutions. Second, the studies of Gervais and collaborators (see 
e.g. [\GERC]) deal with the elliptic sector of the theory. More general 
singular solutions appear in the work [\Mar], but see our remark at the end
of the introduction to Section 5.

\bigskip
\centerline{\bf 6.~Discussion}
\medskip

In this paper  we dealt with the coadjoint orbits of the 
Virasoro algebra and the solution space of the global Liouville equation. 
Our description of the Virasoro orbits is summarized by Figure 1.
We connected our elementary presentation  with the well-known
classifications of the Virasoro orbits [\LP,\Ki,\Wi], whereby
we hope to have made the known results  more easily accessible,
and we also made them more explicit and more complete in some respects.
Our review of the orbits was motivated 
by its subsequent application to the global Liouville equation (1.1).
The main interest of this equation is that its smooth
solutions describe also those solutions of the Liouville equation (1.2) 
that may be singular in terms of $\varphi$  in such a way that 
the Virasoro densities $L_\pm$,
the components of the conformally improved Liouville 
stress-energy tensor, remain smooth.
We have listed all smooth solutions of the global Liouville
equation up to conformal transformations under periodic boundary
condition.
In this way we established the correlation between the 
topological type (given by the number of zeros of $Q$) of the
solutions and their Virasoro orbit type.
The energy (5.1) (with $\kappa >0$)  turned out to be bounded from below 
only in the trivial topological sector that contains
the solutions with Virasoro orbit type  
${\cal B}_0(b) \times {\cal B}_0(b)$.

The antiperiodic, $Q(x^++2\pi, x^- - 2\pi)=-Q(x^+,x^-)$, solutions
of (1.1) can be analyzed analogously to the periodic case, since 
$L_\pm$ are periodic for these solutions, too.
In the antiperiodic case $Q$ has an odd number of zeros 
for $0 \leq \sigma < 4\pi$, and the possible pairings of
the chiral and antichiral Virasoro densities are governed by the
monodromy condition $M_{\Psi_-}=-(M_{\Psi_+})^T$ that replaces (1.7).
Similar to (5.5), we still have $N=(\nu_+ + \nu_-)$ for elliptic
solutions. 
An interesting consequence is that in the $N=1$ 
sector we can pair the orbits as 
${\cal C}(\nu_+) \times {\cal C}(\nu_-)$ for $0<\nu_+ <1$,
$\nu_-=(1-\nu_+)$, 
and therefore the energy is bounded from below on the set of
the corresponding solutions.
However, taking the other monodromies into account, the energy is
not bounded from below in any of the topological sectors
in the antiperiodic case (cf. Ref.~[\Mar]).
For instance, if $N=1$ then the Virasoro orbit type 
${\cal B}_0(b) \times {\cal B}_1(b)$,
on which the energy is unbounded, 
 is also permitted, and any
two solutions with the same $N$ can be connected to each other
along a path of solutions.

Throughout this paper we  required that the Virasoro densities $L_\pm$,
and therefore also the global Liouville field $Q$ and the conformal 
transformations $\alpha_\pm \in \widetilde {\rm Diff}_0(S^1)$,
be given by infinitely many times differentiable functions of $x^\pm$.
{}From  the point of view of the differential  equations (1.1) and (1.3),
it might be natural to relax this stringent requirement and assume instead 
that the conformal transformations and the global Liouville field are thrice 
and twice continuously differentiable functions, 
respectively, while the Virasoro densities are assumed to be only continuous.
Under these requirements  the set of possible 
Virasoro densities becomes larger, 
but so does the group of conformal transformations too.
As a result, the  
classification of the Hill's equations  in Section 3 
and the solutions of the global Liouville equation in Section 5, 
including  
the list of nonequivalent representatives given in the respective sections, 
remain valid without change.

The previous studies [\GPP,\Mar]
of  the singular solutions of the Liouville equation
were already mentioned in the Introduction.
In Ref~[\GPP], whose  analytic methods are rather different from our
group theoretic approach, the admissible singularities of the field 
$\varphi$ and the boundary conditions at infinity 
are specified directly in such a way that their nature is preserved
by the time development, and the regularity of the
stress-energy tensor as well as the existence of certain associated
Noether charges are ensured.
For any fixed solution $\varphi(\tau,\sigma)$, the time development of the
singularities is then interpreted [\GPP] as the dynamics of certain 
`particles' on the line interacting  with each other through a 
velocity dependent potential.
For some particular solutions, the dynamics of the singularities
is even Poincar\'e invariant.
It could be interesting to generalize this picture and describe 
the dynamics of the zeros of our smooth and periodic $Q$ 
subject to (1.1) 
in terms of some system of interacting particles on the circle. 
Some results in Ref.~[\GJ] may be related to this problem.

As we saw  in Section 2, 
the global Liouville system is a 
reduced WZNW model [\Bal] for the group $SL(2,\R)$.
For other groups, 
the reduced WZNW systems [\Bal] contain the standard Toda field theories 
in their trivial topological sector.
Some nontrivial aspects of these  reduced systems have been 
investigated in the point particle case [\FT,\Ful,\KT,\RY].
In the field theoretic case the $SL(3,\R)$ model was studied [\BPT]
using the known description of the symplectic leaves of the
$W_3$-algebra [\KS] that replaces the Virasoro algebra as one goes from 
$SL(2,\R)$ to $SL(3,\R)$.
However, the results are far from complete even  in this 
particular case, and exploring the solution
space of the `global Toda models' is a problem for the future.

Finally, the most interesting problem is of course to better  
understand the role of the singular solutions of the Liouville
and Toda equations at the quantum mechanical level.
As was already mentioned, results in this direction can be found in 
Ref.~[\Mar] and in the works [\GERC]. However, a full treatment including 
all types of singular solutions appears to be still missing.

\bigskip
\medskip
\noindent
{\bf Acknowledgements.}
L.F.\ wishes to thank I.\ Tsutsui for discussions and is grateful to the 
Alexander von Humboldt Foundation for a fellowship. 
This work was supported in part by the Hungarian National Science Fund
(OTKA) under T016251 and T019917.

\medskip
\bigskip
\centerline 
{\bf Appendix A: The number of zeros of $Q$ as a winding number}
\medskip

In this appendix we study the connection between the number of zeros 
$N[Q]$ of the  field $Q$, and the winding number $W[g]$ of the 
corresponding $SL(2,{\bf R})$ valued WZNW field $g$.
Both quantities are topological invariants, so it is natural to assume 
that they are related, and we shall find that the relationship is given 
by eq.~(A.6) below.

We need a concrete description of the  
winding number associated with the   
WZNW field $g(\sigma)\in SL(2,\R)$.
For this we parametrize the $SL(2,\R)$ matrix elements  
$$
g=\pmatrix{a&b\cr c&d\cr}\qquad\qquad ad-bc=1
\eqno(\rA.1)
$$
using the three-dimensional system of coordinates $(\phi,x,y)$ as
$$\eqalign{
d-a=2x\qquad\qquad\qquad d+a&=2r\,{\rm sin}\,\phi\,,\cr
b+c=2y\,\qquad\qquad\qquad b-c&=2r\,{\rm cos}\,\phi\,,\cr}
\eqno(\rA.2)
$$
where $r=\sqrt{1+x^2+y^2}$. We introduced this system of coordinates
because it makes transparent that the group $SL(2,{\bf R})$ has the
topology of $U(1)\times{\bf R}^2$. The winding number  of the WZNW
field $g(\sigma)$ is the number of times $\exp(i\phi(\sigma))$ goes around 
the $U(1)$
circle as the field completes a full period 
$4\pi$ in $\sigma$.
It may be  determined as the regular integral
$$
W[g]={1\over2\pi}\,\int_0^{4\pi}\, d\sigma\,{(b-c)(a^\prime+d^\prime)+
(c^\prime-b^\prime)(a+d)\over4+(a-d)^2+(b+c)^2}\,,
\eqno(\rA.3)
$$
but it is more useful to determine  it directly 
as ${1\over2\pi}$-times the total change in the
angle 
$$
\phi(\sigma)={\rm arc tg}\,\left({a+d\over b-c}\right)(\sigma) 
\eqno(\rA.4)
$$
as $\sigma$ varies over a $4\pi$-period. 

Now we restrict the  WZNW field $g(\sigma)$ to the gauge in (2.8),
where by (2.9) and (2.10) we  have 
$$\eqalign{
a+d&=\ell_+g_{22}-2\kappa P_+^\prime+g_{22}=
g_{22}+{1\over g_{22}}(1+P_+^2-2\kappa P_+g_{22}^\prime)\,,\cr
b-c&=2\kappa  g_{22}^\prime,\,
\qquad (Q=\kappa g_{22}).\cr}
\eqno(\rA.5)
$$
This shows that for the gauge fixed configuration 
$(a+b)/(b-c)$ in (A.4) is singular precisely 
at the points where $Q'$ vanishes, and that 
the sign of
the argument of the ${\rm arc tg}$ function close to a  point where it is
singular is equal to 
${\rm sign\,}(\kappa) {\rm sign}\,(Q)\,{\rm sign}\,(Q^\prime)$.
Clearly,  the contribution to the winding number of an interval between
two consecutive zeros of $Q^\prime$ depends only on the product of these
signs at the beginning and at the end of the interval. Since the
sign of $Q^\prime$ does not change within such an interval, there is
no contribution coming from the interval if the sign of $Q$ does not
change either, as for example between points 1 and 2 or 2 and 3
on Figure 3.

\vskip-2truecm
\hskip1.0truecm
\epsfbox{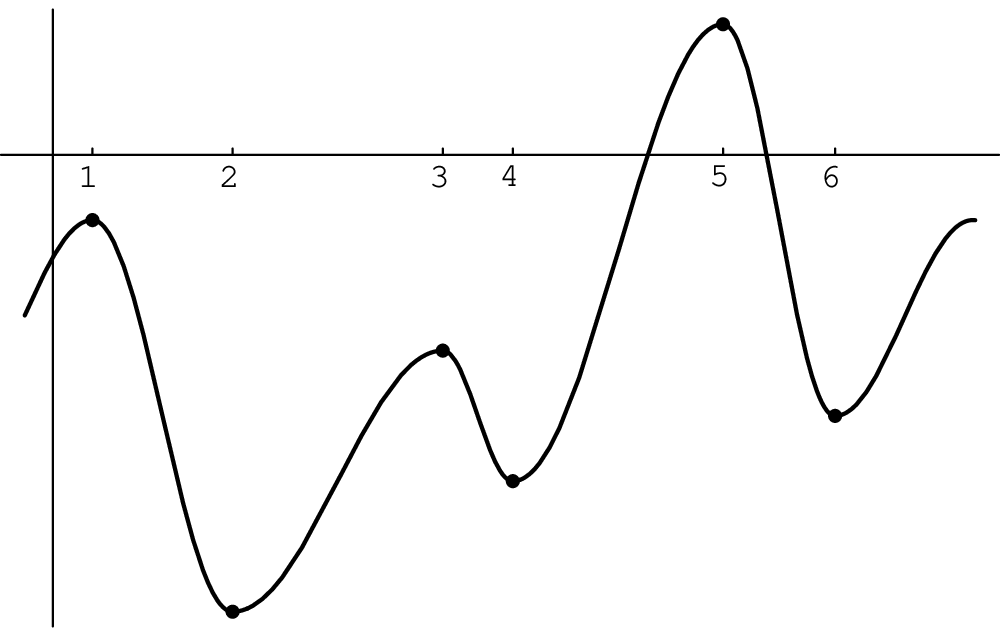}

\vskip-2truecm
\centerline{Figure 3:  \ \ An example of the function $Q(\sigma)$}
\bigskip
\noindent
There is a nonvanishing contribution to $W$ only if $Q$ changes sign
between two consecutive zeros of $Q^\prime$.
This happens  if the interval contains
a zero of $Q$,  and in such a case the contribution is always 
${\rm sign\,}(\kappa) /2$,
independently of whether the curve reaches the zero from below or from
above. (See the interval between points 4 and 5 and the interval
between 5 and 6  on Figure 3.) 
It follows from the above observations  that 
the winding number of a WZNW field $g$ in (2.8) 
corresponding to a 
global Liouville field $Q=\kappa g_{22}$ with $2n$ zeros is 
${\rm sign\,}(\kappa ) n $.
That is we have  
$$
N[Q] = 2\, {\rm sign\,}(\kappa) W[g]
\eqno(\rA.6)$$
for any configuration in (2.8).
This means in particular that  the constraints  in (2.6) 
exclude those WZNW configurations 
with nonzero winding number for which the sign of the winding number 
does not agree with the sign of  $\kappa$.

\bigskip
\centerline{\bf Appendix B:
The little groups for orbits with constant Virasoro densities}
\medskip

We below investigate the little groups for the 
orbits that contain a constant  Virasoro
density, $L = \Lambda \in {\bf R}$, by integrating the equation
$$
\Lambda={\alpha'}^2 \Lambda + S(\alpha).
\eqno({\rm B}.1)$$
Like in Subsection 4.1,  it proves useful to introduce the function 
$$
q= {1\over (\alpha^{-1})'}=\alpha'\circ \alpha^{-1}.
\eqno({\rm B}.2)$$
Recall from (4.9) that $q$ is smooth, $2\pi$-periodic and  positive.

We consider first the orbits with elliptic monodromy,
for which $\Lambda=-{\nu^2\over4}$.
In this case (B.1) reads in terms of $q$ as
$$
-{\nu^2\over4}=-{\nu^2\over 4}q^2-{1\over2}qq''+{1\over4}(q')^2.
\eqno({\rm B}.3)
$$
By the substitution $q=f^2/2$,  this  simplifies to a 
problem of classical mechanics in one dimension:
$$
f''=\nu^2\left({1\over f^3}-{1\over4}f\right),
\eqno({\rm B}.4)
$$
which can be solved by quadrature. The solution will be real if 
the constant of integration, $a$, arising in the first integration,
$$
{1\over2}(f')^2=a-{\nu^2\over2}\left({1\over f^2}+{f^2\over4}\right),
\eqno({\rm B}.5)
$$
satisfies $a\ge {\nu^2\over2}$. In this case the second integration 
yields
$$
q={2a\over\nu^2}+\sqrt{\left({2a\over\nu^2}\right)^2-1}\ {\rm sin}(\nu x+c),
\eqno({\rm B}.6)$$
where $c$ is the corresponding  constant of integration. 
The function $q$ given by (B.6)
is positive for all values of $a$ and $\nu$, and  it is 
$2\pi$-periodic  if and only if 
$$
{\rm (i)}\quad \nu\not\in {\bf N}, \ a={\nu^2\over2};\qquad {\rm (ii)}\quad 
\nu\in {\bf N}, \ a\ge {\nu^2\over2}.
$$
In case (i) one easily gets from the definition of $q$ that $\alpha(x)=
x+\alpha_0$, describing the lift of
the group of rigid rotations 
${\rm Rot}(S^1)=S^1$ 
 to the translation group of ${\bf R}$. In case (ii), which corresponds to 
 orbits with $E_\eta$-type monodromy,   
  we find from 
inverting the relation between $q$ and $\alpha$ that
$$
\alpha (x)={2\over n}{\rm arc tg}\left( {{\rm tg}({n\over2}x+\hat c)\over 
{2a\over n^2}+\sqrt{\left({2a\over n^2}\right)^2-1}}\right) +\alpha_0,
\eqno({\rm B}.7)$$
where $\hat c$ is a new constant of integration, and $\alpha_0$ is a 
constant related to $c$. To exhibit the $SL(2,\R)$ ($SU(1,1)$) character of
these conformal transformations we use the well-known relation between the
inverse trigonometric and logarithmic functions,
which allows us to rewrite (B.7) as
$$
\alpha (x)={1\over in}{\rm ln}{ A e^{inx}+\bar B \over B e^{inx}+\bar A},
$$
where
$$
A={M+1\over 2\sqrt{M}}e^{i(2{\hat{c}}+{\alpha_0}n)/2}, \quad
\bar B={M-1\over 2\sqrt{M}}e^{-i(2{\hat{c} }-{\alpha_0}n)/2},\quad
M={2a\over n^2}+\sqrt{\left({2a\over n^2}\right)^2-1}.
$$ 
These equations show that
$\chi(\alpha)\in {\rm Diff_0}(S^1)$,
see eq.~(3.4),  operates 
on $z=e^{ix}\in S^1$ according to 
$$
\chi(\alpha) : z\mapsto 
\left[ {Az^n+\bar B\over Bz^n+\bar A}\right]^{1\over n} 
\qquad\hbox{where}\qquad 
\pmatrix{A&\bar B\cr  B&\bar A\cr}\in SU(1,1).
$$
Therefore, for any $n\in {\bf N}$,
 the little group of $L=-{n^2\over 4}$ in 
${\rm Diff}_0(S^1)$ is an $n$-fold cover
of the M\"obius group $PSL(2,\R)$. 

In the case of orbits with hyperbolic monodromy,
for which  $\Lambda=b^2$ with $b>0$ as in (3.52), 
the integration of the equation replacing (B.4),
$$
f''=b^2\left(f-{4\over f^3}\right),
$$
yields 
$$
{1\over2}(f')^2=a+b^2\left({2\over f^2}+{f^2\over2}\right),
\eqno({\rm B}.8)
$$ 
instead of (B.5). The constant $a$ must satisfy $a\ge -2b^2$ 
to guarantee the positivity of the r.h.s. Since the integration of eq.~(B.8) 
involves hyperbolic functions rather than trigonometric ones, it is not 
difficult to see that the only periodic solution is $f^2\equiv 2$, which is  
obtained for $a=-2b^2$. This solution, obviously, leads again to 
$\alpha (x)=x+\alpha_0$. 

Among the orbits with parabolic monodromy only ${\cal P}_0^+$ has a constant 
Virasoro density, and this corresponds to $\Lambda=0$. In this case eq.~(B.4) 
simplifies to $f''=0$,
which is solved by $f=ax+c$. This solution is periodic only for $a=0$, 
and with this choice (after tuning the constant of integration  
$c$ as required by
eq.~(1.10)) it leads again to  $\alpha (x)$ describing the lift of the 
group of rigid rotations.

\bigskip
\centerline{\bf Appendix C:  Construction of conformal transformations 
for ${\cal B}_n(b)$ and ${\cal P}_n^q$}
\medskip

We here show that  any  two solution vectors, 
say $\Psi=\pmatrix{\psi_1 &\psi_2}$ and 
$\overline\Psi=\pmatrix{\overline\psi_1&\overline\psi_2}$,
of Hill's equation that have the same 
monodromy matrix $B_\eta(b)$ (or $P_\eta^q$) 
and discrete invariant $n>0$ (3.43) can be transformed into each other 
by a conformal transformation.
In particular,  this proves that $\Psi$ in (3.58) 
(and that in (3.71)) is indeed a representative for the solutions
associated with the Virasoro orbit ${\cal B}_n(b)$ (and respectively 
with ${\cal P}_n^q$).

For definiteness, let us concentrate on the case ${\cal B}_n(b)$. 
Denote the zeros of the component $\psi_2$ (the number of
which, within a $2\pi$ period, is $n$) by $x_k$ and arrange the indexing
of these zeros so that
$$\eqalign{
0&\leq x_1<x_2 \, \dots\, x_n<2\pi\,,\cr 
x_{k+mn}&=x_k+2m\pi\,,\qquad\qquad k=1,\dots,n\ \ \ 
\ \ \ \ m\in {\bf Z}\,.\cr}
\eqno({\rm C}.1)
$$ 
Because of the Wronskian condition, the derivative of $\psi_2$ cannot
vanish at the points $x_k$ and we have
$$\eqalign{
\psi_2^\prime(x_k)&=r_k\not=0\,,\cr
\psi_1(x_k)&={-1\over r_k}\,.\cr}
\eqno({\rm C}.2)
$$
The ratio
$$
u(x)={\psi_1(x)\over\psi_2(x)}\qquad\qquad x\not=x_k
\eqno({\rm C}.3)
$$
is well-defined and smooth away from the set of zeros and has the
properties
$$\eqalignno{
u(x+2\pi)&=e^{4\pi b}u(x)\,,&({\rm C}.4{\rm a})\cr
u^\prime(x)&={1\over\psi_2^2(x)}\,>0\,.&({\rm C}.4{\rm b})\cr}
$$
Near the point $x_k$ $u(x)$ can be written as
$$
u(x)\,=\,{-1\over (x-x_k) r_k^2 }+\beta_k(x)\,,
\eqno({\rm C}.5)
$$
where $\beta_k(x)$ is smooth around $x_k$.
Using (C.4) and (C.5) we see that between two consecutive zeros
$x_k$ and $x_{k+1}$
$u(x)$ increases monotonically from $-\infty$ to $+\infty$.
This allows us to define a set of functions $v_k$, the
inverses of $u$, by
$$
v_k(u(x))=x\qquad\qquad x_k<x<x_{k+1}\,.
\eqno({\rm C}.6)
$$
Note that $v_k(y)$ is defined for all $y$, it is smooth and
monotonically increases from $x_k$ to $x_{k+1}$ as $y$ varies
from $-\infty$ to $+\infty$. From (C.6) and the properties of the
solution of Hill's equation it also follows that
$$
u(v_k(y))=y\qquad\qquad -\infty<y<\infty
\eqno({\rm C}.7)
$$
and
$$
v_{k+n}(e^{4\pi b}y)=v_k(y)+2\pi\,.
\eqno({\rm C}.8)
$$
We are now able to define the element
$\alpha \in \widetilde{\rm Diff}_0(S^1)$ that transforms
$\overline\Psi$ to $\Psi$:
$$\eqalign{
&\alpha(x_k)=\overline x_{k+d}\qquad\qquad\qquad\qquad k\in{\bf Z}\,,\cr
&\alpha(x)=\overline v_{k+d}(u(x))\qquad\ \ \ \ \ \ \ \ 
x_k<x<x_{k+1}\,.\cr}
\eqno({\rm C}.9)
$$
Here (and in the following) we use the convention that overlined 
objects 
are defined for $\overline\Psi$ analogously to the corresponding 
quantities for $\Psi$. Note the shift $k\rightarrow k+d$ in the
index of the zeros. This shift has to be chosen so that the 
signs of
$r_k$ and $\overline r_{k+d}$ coincide,
which is clearly necessary for $\alpha$ to transform 
$\overline \Psi$  to $\Psi$.

Using (C.8) it is easy to show that $\alpha(x)$ as defined by (C.9) 
satisfies $\alpha(x+2\pi)=\alpha(x)+2\pi$ and
is smooth and strictly monotonically increasing for all $x\not=x_k$.
To show that it is a proper conformal transformation, we still have to
show that it is also differentiable at the zeros $x_k$.
 
Before we do that we differentiate (C.7) and use (C.4b) to obtain
$$
v_k^\prime(y)={1\over u^\prime(v_k(y))}=\psi^2_2(v_k(y))\,.
\eqno({\rm C}.10)
$$
Using this (for the analogous overlined objects) and taking the
derivative of the definition (C.9) we get
$$
\alpha^\prime(x)=\overline\psi_2^2(\overline v_{k+d}(u(x)))
{1\over\psi_2^2(x)}={\overline\psi_2^2(\alpha(x))\over\psi_2^2(x)}\,.
\eqno({\rm C}.11)
$$
Finally, on account of the equality of the sign of $r_k$ and 
$\overline r_{k+d}$, 
we  can take the square root as
$$
{1\over\sqrt{\alpha^\prime(x)}}\overline\psi_2(\alpha(x))=
\psi_2(x)
\qquad\qquad x\not=x_k\,,
\eqno({\rm C}.12)
$$ 
which is the  required conformal transformation from 
$\overline\Psi$ to $\Psi$.
More precisely, we still have to show that $\alpha$ is smooth and (C.12)
holds also at the set of zeros $x_k$. We show this by first observing
that (C.9) and (C.7) imply
$$
\overline u(\alpha(x))=\overline u(\overline v_{k+d}(u(x)))=u(x)\,,
\eqno({\rm C}.13)
$$
which, when multiplied by (C.12), gives
$$
{1\over\sqrt{\alpha^\prime(x)}}\overline\psi_1(\alpha(x))
=\psi_1(x)\,.
\eqno({\rm C}.14)
$$ 
By continuity, (C.14) holds for all $x$ (including the set of zeros),
and this allows us to express $\alpha^\prime$ as
$$
\alpha^\prime(x)={\overline\psi_1^2(\alpha(x))\over\psi_1^2(x)}, 
\eqno({\rm C}.15)
$$
showing that
$\alpha(x)$ is indeed smooth also around the points $x_k$.

To summarize, we have proved that any solution vector 
of Hill's equation in the class ${\cal B}_n(b)$, $n>0$
with monodromy matrix $B_\eta(b)$ 
can be expressed as the conformal transform
of some unique representant, which  can be chosen, for
example, as in (3.58). 
Incidentally, the above consideration also clarifies the structure of 
the little group for the Virasoro orbit ${\cal B}_n(b)$.
Because of the structure of $G[B_\eta(b)]$ in (3.22b),
we see that $\alpha\in \widetilde{\rm Diff}_0(S^1)$ 
belongs to the little group $\tilde G[\overline L]$ of 
$\overline L={\overline \psi_2'' /\overline \psi_2}$
if and only if it transforms $\overline \psi_2$ into itself up to 
a nonzero constant rescaling. 
The sign of this rescaling and the parity of the index
shift $d$ in (C.9) are related and therefore the elements of 
$\tilde G[\overline L]$ can be 
labelled by the shift $d$ and the absolute value, $\lambda$, 
of the rescaling parameter. 
Since we can construct a unique element 
$\alpha_{\lambda, d}\in \tilde G[\overline L]$ for any 
$\lambda \in {\bf R}_+$ and $d\in {\bf Z}$,
the group 
$\tilde G[\overline  L]$ is isomorphic to ${\bf R}_+ \times {\bf Z}$,
where ${\bf R}_+$ is the multiplicative group of the positive real
numbers.
Factoring by the group of translations by multiples of $2\pi$,
whose elements are $\alpha_{\lambda,mn}$ with 
$\lambda=e^{-2\pi bm }$
for $m\in {\bf Z}$,
we obtain that $G[\overline L]\subset {\rm Diff}_0(S^1)$ is isomorphic
to ${\bf R}_+\times {\bf Z}_n$.

Our considerations are also valid in the parabolic 
case ${\cal P}_n^q$ if we replace (C.4a) by
$$
u(x+2\pi)=u(x)+q
\eqno({\rm C}.4{\rm a}^\prime)
$$
and (C.8) by
$$
v_{k+n}(y+q)=v_k(y)+2\pi\,.
\eqno({\rm C}.8^\prime)
$$
For any $L$ on the Virasoro orbit ${\cal P}_n^q$ with $n>0$, 
the little groups $\tilde G[L]$ and $G[L]$
are again isomorphic to ${\bf R}_+\times {\bf Z}$ and to 
${\bf R}_+\times {\bf Z}_n$, respectively.
\vfill\eject

\bigskip
\centerline{\bf Appendix D: The connectedness of the phase space with
$2n$ zeros}
\medskip
In this Appendix we show that beyond the number of zeros, which is
a topological invariant as discussed in Appendix A, there are no
additional topological invariants in the solution space of the global
Liouville equation. In other words, any two solutions of the global 
Liouville equation with the same number of zeros can be continuously 
connected along a path of solutions. (In the case of no zeros  the two
solutions must also be of the same sign).

To show this we first note that any solution can be continuously deformed
into one of the representatives listed in Section 5. Indeed, any solution
is a conformal transform of one of the representatives and the conformal
group as defined in (1.10) is connected. For the case of no zeros the
representatives are given by (5.11) and it is clear that apart from 
the overall sign they are all connected (by changing $b$ along 
the first vertical line of Figure 1).

In the case of $N=2n\not=0$  zeros there are representatives with
elliptic, $E$-type, hyperbolic and parabolic monodromy. It is clear that
the elliptic representatives (given by (5.4)) together with the
$E$-type representatives of (5.7) are connected (by changing
$\nu_+=N-\nu_-$ along the horizontal line of Figure 1).
On the other hand, the hyperbolic and parabolic representatives 
with $N=n_++n_-$ where neither $n_+$ nor $n_-$ vanishes are easily connected
to the corresponding $E$-type representatives. For the hyperbolic
case the connection is along the vertical lines of Figure 1, since the
representatives (5.16) are simply reduced to (5.7) in the limit 
$b\rightarrow0$. For the parabolic case of (5.20) we have to move along
the curve (3.77) parametrized by $a$ from $a=1$ to $a=\infty$, i.e. from
(5.20) to (5.7). 

Hyperbolic and parabolic solutions with $n_+=0$ (and the
analogous $n_-=0$ cases) are more difficult to connect with the
elliptic ones and here we need a different way of connecting the
solutions. (The reason why a simple connection along Figure 1 is not
available for these cases is the existence of the empty circle at
$n=0$ which blocks the way between the first vertical line and the
horizontal line of the figure.)

We will make use of a different way of representing the solutions of the
global Liouville equation, namely, by considering it to be an initial-value
problem. It is clear that a solution can be uniquely represented by its
value and first time derivative at $\tau=0$:
$$
q(\sigma)=Q(0,\sigma)\qquad{\rm and}\qquad
k(\sigma)={\partial\over\partial\tau}\, Q(0,\sigma)\,.
\eqno({\rm D}.1)
$$
It would be possible to connect any two solutions, $\{q_1,k_1\}$ 
and $\{q_2,k_2\}$ along a straight line:
$$\eqalign{
q(\sigma,t)&=(1-t)\,q_1(\sigma)+t\,q_2(\sigma)\cr
k(\sigma,t)&=(1-t)\,k_1(\sigma)+t\,k_2(\sigma)\cr}
\eqno({\rm D}.2)
$$
for $t$ between 0 and 1 if it did not follow from the global
Liouville equation (1.1) that at all points where $q(\sigma)$ vanishes,
$$
(q^\prime)^2=1+k^2
\eqno({\rm D}.3)
$$
must hold. However, it is clear that the pair of functions (D.2) does not 
satisfy (D.3) in general (only for $t=0,1$) and therefore (D.2) does not 
represent a curve in the space of solutions.

If, on the other hand, the two solutions are such that the set of zeros, 
$\{x_1,...,x_{2n}\}$ is the same for $q_1$ and $q_2$ and at each zero 
$$
q^\prime_1=q^\prime_2\qquad{\rm and}\qquad k_1=k_2\,,
\eqno({\rm D}.4)
$$
hold, then (D.2) can after all be used to connect the two solutions 
in the space of solutions.

Let us consider now the solutions (5.15). The corresponding pair, 
$\{q,k\}$, has the following properties: $q(\sigma)$ has
$2n$ zeros at equal ($2\pi/n$) spacing. At each of the zeros we have
$$\eqalign{
{\rm sign}\,(q^\prime k)&=-1\,,\cr
|q^\prime|&={f+1\over\sqrt{2f+1}}\,,\cr 
|k|&={f\over\sqrt{2f+1}}\,,\cr}
\eqno({\rm D}.5)
$$
where
$$
f={n-{3b\over2}+{b^2\over8n}\over2b}\,.
\eqno({\rm D}.6)
$$
 
The $\{q,k\}$ pair corresponding to the parabolic solution (5.19) has
similar properties, namely, $q$ has also $2n$ equally spaced zeros and
(D.5) holds at the zeros with
$$
f=\pi-1\,.
\eqno({\rm D}.7)
$$

Finally, for the elliptic case, using (5.4), we find that at the similarly 
equally spaced zeros (D.5) holds with
$$
f={n\over\nu_+}-1\,.
\eqno({\rm D}.8)
$$

Thus, the parabolic and the hyperbolic solutions can be connected with the
elliptic ones using (D.2) if for the latter we choose
$$
\nu_+={n\over\pi}\qquad{\rm and}
\qquad\nu_+={2bn\over n+{b\over2}+{b^2\over8n}}\,,
\eqno({\rm D}.9)
$$
respectively.

\vfill\eject

\centerline{\bf References}

\parskip=3pt
\medskip 

\item{[\Sei]}  
N.\ Seiberg, {\sl Prog.\ Theor.\ Phys.\ Suppl.\ } {\bf 102} (1990) 319.

\item{[\GPP]} 
G.P.\ Dzhordzhadze, A.K.\ Pogrebkov and M.C.\ Polivanov,
{\sl Theor.\ Math.\ Phys.\ }{\bf 40}(2) (1980) 706;  
A.K.\ Pogrebkov and M.K.\ Polivanov,
{\sl Sov.\ Sci.\ Rev.\ C (Math.\ Phys.)} Vol.~5 (1985) pp.\ 197-272,
and references therein.

\item{[\Mar]}
L.Johansson, A. Kihlberg and R. Marnelius, {\sl Phys. Rev.} {\bf D29} 
(1984) 2798;
L. Johannson and R. Marnelius, {\sl Nucl. Phys.} {\bf B254} (1985) 201.

\item{[\Bal]}
J.\ Balog, L.\ Feh\'er, P.\ Forg\'acs, L.\ O'Raifeartaigh and 
A.\ Wipf,  {\sl Phys. Lett.} {\bf B227} (1989) 214;
{\sl Ann. Phys. (N.Y.)} {\bf 203} (1990) 76.

\item{[\WZ]}  
E. Witten,
{\sl Commun. Math. Phys.} {\bf 92} (1984) 483.

\item{[\FT]}
L.\ Feh\'er and I.\ Tsutsui, 
{\sl Prog.\ Theor.\ Phys.\ Suppl.\ } {\bf 118} (1995) 173; 
{\sl J. Geom. Phys.} {\bf 21} (1997) 97.

\item{[\MW]}
W.\ Magnus and S.\ Winkler,
Hill's Equation, Interscience, 1966 (Dover, 1979).

\item{[\LP]} 
V.P.\ Lazutkin and T.F.\ Pankratova,
{\sl Funct.\ Anal.\ Appl.\ } {\bf 9}(4) (1975) 41.

\item{[\Ki]}
A.A.\ Kirillov, 
{\sl Funct.\ Anal.\ Appl.\ } {\bf 15}(2)  (1981) 135;
Springer Lecture Notes in Mathematics, vol.\ 970 (1982) 101.

\item{[\Seg]} 
G.\ Segal, 
{\sl Commun.\ Math.\ Phys.\ } {\bf 80} (1981) 301.

\item{[\Wi]} 
E. Witten,
{\sl Commun. Math. Phys.} {\bf 114} (1988) 1.

\item{[\OK]}
V.Yu.\ Ovsienko and B.A.\ Khesin, 
{\sl Funct.\ Anal.\ Appl.\ } {\bf 24}(1) (1990) 33.

\item{[\GRS]}
A.\ Gorsky, B.\ Roy and K.\ Selivanov,
{\sl JETP Lett.\ } {\bf 53}(1) (1991) 64.

\item{[\G]}
L.\ Guieu, 
Sur la geometrie des orbites de la representation coadjointe du
groupe de Bott-Virasoro, PhD thesis, Universite d'Aix-Marseille I, 1994.

\item{[\Fad]}
L.D.\ Faddeev and L.\ Takhtajan
Springer Lecture Notes in Physics, vol.\ 246 (1986) 166.

\item{[\Ald]}
E.\ Aldrovandi, L.\ Bonora, V.\ Bonservizi and R.\ Paunov,
{\sl Int.\ J.\ Mod.\ Phys.\ } {\bf A9} (1994) 57.

\item{[\Liu]}
J.\ Liouville, 
{\sl J.\ Math.\ Pures Appl.\ } {\bf 18} (1853) 71.

\item{[\THE]}
E. Braaten, T. Curtright and C. Thorn, 
{\sl Phys.\  Lett.\ } {\bf B118} (1982) 115; {\sl Ann.\ Phys.\ (N.Y.)} 
{\bf 147} (1983) 365; T. Curtright and C. Thorn, {\sl Phys.\ Rev.\ Lett.\ }
{\bf 48} (1982) 1309.

\item{[\OTW]} H.J. Otto and G. Weigt {\sl Phys.\ Lett.\ } {\bf B159} (1985) 
341; {\sl Z.\ Phys.\ } {\bf C31} (1986) 219.

\item{[\GERC]} J.L. Gervais and A. Neveu, {\sl Nucl.\ Phys.\ } {\bf B199} 
(1982) 59; ibid {\bf B209} (1982) 125; ibid {\bf B224} (1983) 329; J.L. Gervais 
and J. Schnittger, {\sl Nucl.\ Phys.\ } {\bf B413} (1994) 433.

\item{[\GJ]}  
A.\ Gorsky and A.\ Johansen,
{\sl Int.\ J.\ Mod.\ Phys.\ } {\bf A10} (1995) 785.

\item{[\Ful]}
T.\ F\"ul\"op,
{\sl J.\ Math.\ Phys.\ } {\bf 37} (1996) 1617.

\item{[\KT]}
H.\ Kobayashi and I.\ Tsutsui,
{\sl Nucl.\ Phys.\ } {\bf B472} (1996) 409.

\item{[\RY]}
A.V.\ Razumov and V.I.\ Yasnov,
Hamiltonian reduction of free particle motion on group $SL(2,{\bf R})$, 
preprint hep-th/9609030. 

\item{[\BPT]}
Z.\ Bajnok, L.\ Palla and G.\ Tak\'acs, 
{\sl Nucl.\ Phys.} {\bf B385} (1992) 329.

\item{[\KS]}
B.A.\ Khesin and B.Z.\ Shapiro,   
{\sl Commun.\ Math.\ Phys.\ } {\bf 145} (1992) 357.

\bye